\documentclass[12pt]{article}
\RequirePackage[OT1]{fontenc}
\usepackage{amsmath,amssymb,amsthm}
\usepackage{natbib}
  \bibpunct{(}{)}{;}{a}{}{,} 
\usepackage{graphicx}
\usepackage{bm}
\usepackage{caption} 
\usepackage{xcolor}

\usepackage[vlined]{algorithm2e}
\usepackage[colorlinks=true,linkcolor=blue,citecolor=blue,urlcolor=blue,bookmarks=false]{hyperref}
\usepackage{verbatim}
\usepackage[top=1in, bottom=1in, left=1in, right=1in]{geometry}
\usepackage{setspace}
\usepackage{array}
\usepackage{rotating}
\usepackage{float}
\RequirePackage{xspace}
\RequirePackage{tabularx}
\RequirePackage{comment}

\usepackage{fancyhdr}
\DeclareMathOperator{\Var}{Var}
\DeclareMathOperator{\Cov}{Cov}

\DeclareMathOperator{\Tr}{tr}

\newtheorem{theorem}{Theorem}
\newtheorem{lemma}{Lemma}
\newtheorem{proposition}{Proposition}

\newtheorem{definition}{Definition}
\newtheorem{assumption}{Assumption}

\begin{document}

\pagestyle{empty}

\title{Causal inference with bipartite designs: \\ A generalized propensity score approach\thanks{We would like to thank Daniel Saban\'es Bov\'e, Kay Brodersen, Tom Ferris, Guido Imbens, Sebastien Lahaie, Georgia Papadogeorgou, Lewis Rendell, and Corwin Zigler for valuable comments and suggestions. All remaining errors are our own.}}

\author{
Nick Doudchenko
\and Minzhengxiong Zhang
\and Yao Zhao
\and Evgeni Drynkin
\and Edoardo M. Airoldi
\and Vahab Mirrokni
\and Jean Pouget-Abadie
}

\date{}

\maketitle
\thispagestyle{empty}

\begin{abstract}
Bipartite experiments—in which one set of units receives a treatment while outcomes are measured on another set—have recently garnered attention for their ability to capture interference across two distinct populations, such as buyers and sellers in online marketplaces. However, analyzing these experiments can be challenging, given that exposure is neither purely binary nor independent across units. In this paper, we propose a unified framework for causal inference in bipartite designs that leverages generalized propensity scores (GPS) to estimate exposure-response functions. Under standard unconfoundedness assumptions, we show that our GPS-based estimators are unbiased and derive theoretical bounds on their variance. We further introduce practical modeling and weighting strategies—along with double deconfounding methods—that integrate the GPS into both the outcome model and the assignment mechanism. Through extensive simulations, we demonstrate that these approaches achieve substantial bias reduction compared to naive methods. We also illustrate their effectiveness in a real-world application using an Amazon Pet Supplies dataset, where controlling for network structure proves critical to drawing valid causal conclusions. Our results underscore the importance of bipartite designs in contexts with significant interference and highlight how GPS-based methods can bolster the reliability of causal effect estimates in such settings.

\vfill

\end{abstract}

\newpage
\tableofcontents

\newpage

\pagestyle{fancy}
\setcounter{page}{1}

\section{Introduction}

Most experiments in both academic and industrial contexts assume that the units receiving the treatment and the units whose outcomes are measured are one and the same. Bipartite experiments, however, depart from this assumption. In these experiments—recently studied in the causal inference literature by \citet{papadogeorgou2019causal}, \citet{pouget2019variance}, and \citet{zigler2021bipartite}—two distinct sets of units form a bipartite graph. One set, referred to as the "diversion units," is assigned the treatment, while the other set, the "outcome units," can be affected by this treatment through their exposure to treated units on the opposite side of the bipartite graph.

Consider, for example, a buyer-item marketplace such as Amazon or Airbnb, where the treatment modifies an item's offer (e.g., through price discounts or faster delivery). Randomly assigning treatment to buyers may raise practical concerns: buyers might perceive unfairness if they observe different offers for the same item. Meanwhile, assigning treatment at the item level in a conventional (non-bipartite) setting would entail measuring outcomes at the item level—potentially creating a different challenge. The presence of substitute items, for instance, can violate the stable unit treatment value assumption (SUTVA), a key condition for obtaining unbiased causal effect estimates. As proposed by \citet{papadogeorgou2019causal}, \citet{pouget2019variance}, and \citet{zigler2021bipartite}, a more robust design is to assign treatment at the item level but measure outcomes for buyers.

In bipartite settings, outcome units (buyers, in this example) are not inherently labeled as treated or control. Instead, to obtain causal estimates, practitioners must associate each outcome with a measure of the treatment exposure conveyed via edges in the bipartite graph. Whether this graph is weighted or unweighted, we assume it is fully known and that it dictates the extent of each unit's exposure. In the marketplace example, buyers who predominantly interact with treated items might be deemed "highly exposed," while those who never interact with treated items are "never exposed." The concept of exposure-whether continuous or categorical, scalar- or vector-valued-derives from the bipartite graph and the treatment assignment on the diversion side, making it a random variable that enables causal inference.

In this paper, we focus on estimating causal effects in bipartite designs. We adopt the linear exposure assumption proposed by \citet{pouget2019variance}, although their primary emphasis was on optimizing the variance of commonly used estimators, rather than ensuring unbiased causal inference. Building on ideas from \citet{imbens2000role}, \citet{hirano2004propensity}, and \citet{forastiere2021identification}, we propose a propensity-score-based estimator specifically adapted to bipartite experimental settings. While our approach refines their foundational contributions, notable differences remain-particularly concerning treatment assignments and their independence across units. We establish that our estimator is unbiased under a set of reasonable assumptions for general bipartite graphs, and we discuss key practical considerations for its implementation as well as the resulting statistical inferences.

The remainder of the paper is organized as follows. Section~2 introduces the bipartite experiment framework and the notion of network exposure. In Section~3, we develop unbiased estimators of the exposure-response function based on the generalized propensity score, explore upper and lower bounds on their variance, and discuss considerations for observational data. Section~4 presents estimation algorithms for static and dynamic bipartite networks. In Section~5, we show that standard bootstrap methods yield correct coverage under an uncorrelated error model, and we further demonstrate that our parametric bootstrap approach achieves correct coverage under correlated errors. Section~6 reports simulation studies of bias, efficiency, and coverage. In Section~7, we apply the methodology to the Amazon Pet Supplies data. We conclude in Section~8 with a discussion of directions for future work.

\section{The bipartite experiment framework}

\subsection{The bipartite network}

We refer to the units that receive treatment or control as \textit{diversion units}, and the units for which outcomes of interest are measured as \textit{outcome units}. These two sets of units are distinct and jointly form a bipartite graph consisting of $N$ outcome units and $M$ diversion units. This distinction is critical for constructing more flexible and representative models of treatment response in bipartite settings where interference is common. Each edge $(i, j)$ connecting outcome unit $i \in$ $\{1, \ldots, N\}$ and diversion unit $j \in\{1, \ldots, M\}$ is associated with a known weight $W_{i j} \in \mathbb{R}$, which remains fixed regardless of any treatment assignment. Figure 1 provides a visual illustration of this setup.

\begin{center}
\includegraphics[scale=.5]{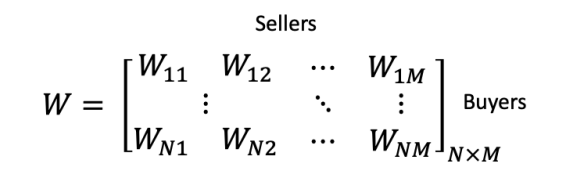}
\captionof{figure}{Bipartite Network Matrix}
\end{center}

The observed outcome of outcome unit $i$ is denoted by $Y_i$, while the treatment assignment for diversion unit $j$ is represented by $Z_j \in\{0,1\}$, where $Z_j=1$ indicates that unit $j$ is treated and $Z_j=0$ indicates it is not treated. 

\subsection{Network exposure}

In Figure 2, we illustrate a simple bipartite network. The treatment exposure $E_i$ for outcome unit $i$ is defined as a function of both the bipartite graph and the treatment assignment $\mathbf{Z}=\left\{Z_j\right\}_{j \in[1, M]}$. Since the graph is assumed to remain fixed, we often write $E_i(\mathbf{Z})$ to denote the treatment exposure of unit $i$ under the assignment $\mathbf{Z} \in\{0,1\}^M$. The precise functional form of this exposure depends on the application and should be specified by domain experts. We assume that it is known, probabilistic, and fully captures all variations in potential outcomes:

$$
\forall i \in[1, N], \quad \forall \mathbf{Z} \in\{0,1\}^M, \quad Y_i(\mathbf{Z})=Y_i\left(E_i(\mathbf{Z})\right)
$$

\begin{center}
\includegraphics[scale=.5]{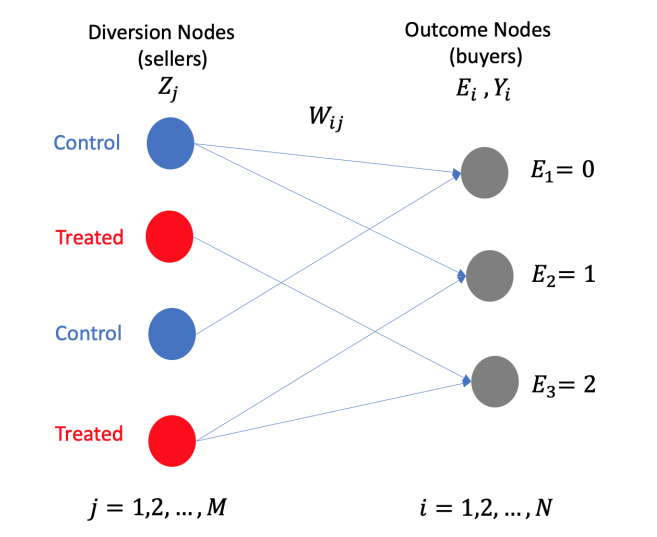}
\captionof{figure}{Bipartite Network Example (Sellers and Buyers)}
\end{center}

In the examples of \citet{papadogeorgou2019causal}, the outcome units are hospitals. The outcome at each hospital is affected by two components: a “direct effect” triggered by the treatment status of the nearest power plant (the diversion unit), and an “indirect effect” determined by the proportion of upwind power plants that are treated. \citet{pouget2019variance} introduce a slightly different exposure function, known as the linear exposure assumption. Under this assumption, the exposure for outcome unit $i$ is the weighted sum of treated neighboring diversion units:

$$
\forall i \in[1, N], \quad E_i(\mathbf{Z})=\sum_{j=1}^M W_{i j} Z_j
$$
Although the results presented in this paper do not rely on a specific exposure function, for simplicity we often adopt the linear exposure assumption. In this formulation, the exposure $E$ is given by $f_E(Z, W)$, a function of the network $\left\{W_{i j}\right\}_{(N, M)}$ and the diversion-side treatment assignment $Z$.

Our main interest lies in the marginal exposure-outcome function $\mu(e)=\mathbb{E}[Y(e)]$ for each $e \in \mathcal{E}$. To construct estimands in a bipartite design, it is useful to consider the exposure-outcome curve, mapping each level of exposure to the population mean of the potential outcome at that exposure level: $\mu: e \mapsto \mathbb{E}\left[Y_i(e)\right]$. When exposure is confined to $[0,1]$-as in the linear exposure setting with appropriately normalized weights-one focal estimand is $\mu(1)-\mu(0)$. This quantity serves as the bipartite-design analogue of the population average treatment effect, gauging the effect of treating all diversion units versus treating none. Another potential estimand of interest is the derivative of the exposure-outcome curve, which represents the marginal impact of a small change in exposure at a given exposure level.

\section{Unbiased estimation with the generalized propensity score}

To obtain accurate estimates based on the example in the previous section, it is crucial to account for the heterogeneity in exposure distributions among outcome units. In this section, we introduce estimators for causal effects, drawing inspiration from the generalized propensity score framework developed for multivalued and continuous treatments \citep{imbens2000role,hirano2004propensity,imai2004causal,robins2000marginal,rosenbaum1983central}. We also build on insights from the literature on interference in experiments \citep{aronow2017estimating,savje2024causal}. We demonstrate that these estimators are unbiased under a specific set of assumptions. We begin by outlining the standard assumptions necessary for our results.

\subsection{Unconfoundedness assumptions}

\setcounter{assumption}{-1}
\begin{assumption}[Fixed network weights]\label{ass:fixedweights}
The graph weights $\left\{W_{i j}\right\}_{N, M}$ must remain unaffected by the treatment assignment $\mathbf{Z}$. Formally,

$$
\mathbf{Z} \perp\left\{W_{i j}\right\}_{N, M}
$$
This requirement means that the vector of treatment assignments $\mathbf{Z}$ is independent of all network weights.
\end{assumption}

For instance, in \citet{papadogeorgou2019causal}, the bipartite graph is defined by the fixed geographic distance between power plants and hospitals, and hence does not depend on whether a power plant is treated. In the market-setting study of \citet{pouget2019variance}, the bipartite graph reflects buyers' item-category preferences, again assumed to be unaffected by treatment. Nevertheless, it is conceivable that a product might become more or less appealing after a discount, potentially violating this assumption.

In general, this assumption states that network connections-such as item rankings, product promotions, or discounts-are formed independently of the actual treatment assignments. Often, a buyer's demand or intent to purchase predates (and is independent of) how the network is formed: for instance, acquiring a pet leads to a subsequent need for pet food, at which point the buyer searches online and creates network ties to various pet-food items. However, in scenarios like Amazon Prime Day, the treatment (i.e., large discounts on many products) itself might alter how buyers interact with items, making it unclear whether the discount drives the formation of new connections or vice versa. In such cases, Assumption 0 may be violated.

In many observational studies involving network data, the response variable $Y_i$ is closely related to the network weights $W_{i j}$. For example, in a seller-buyer network, browsing multiple product pages can build a buyer's confidence in making a purchase, and in social networks, having more friends often correlates with higher engagement and content creation. These relationships illustrate how deeply network connections can be entwined with outcomes and thus emphasize the importance of isolating the effect of the exposure $E_i$. 

\begin{assumption}[Strong unconfoundedness]\label{ass:strong}
Conditioned on the bipartite graph weights $\mathbf{W}=\left(W_1, \ldots, W_N\right)^T$, where $W_i=$ $\left(W_{i 1}, \ldots, W_{i M}\right)^T$, the observed exposure $E_i$ for outcome unit $i$ is independent of all its potential outcomes over the entire range of exposure levels:

$$
E_i \perp\left\{Y_i(e): e \in[0,1]\right\} \mid \mathbf{W}
$$
\end{assumption}

In other words, once we condition on $\mathbf{W}$, knowing the realized exposure $E_i$ provides no further information about any of the potential outcomes $Y_i(e)$.

\begin{assumption}[Weak unconfoundedness]\label{ass:weak}
A weaker requirement states that for each specific exposure level $e$, the potential outcome $Y_i(e)$ is conditionally independent of the realized exposure $E_i$, given the graph weights:

$$
Y_i(e) \perp E_i \mid \mathbf{W}, \quad \text { for each } e \in[0,1]
$$
\end{assumption}

Thus, for each $e$, once $\mathbf{W}$ is known, the realized exposure does not reveal any additional information about $Y_i(e)$.

In many real-world scenarios-particularly those with continuous treatments-strong and weak unconfoundedness tend to coincide. It can be difficult to imagine an applied setting in which one holds but not the other, though it is possible to construct theoretical counterexamples in measure-theoretic or pathological contexts. In practical bipartite experiments, these assumptions usually operate similarly. For instance, in the setting described by \citet{pouget2019variance}, if the treatment assignment is Bernoulli or completely randomized and the linear treatment exposure assumption holds, then -conditioning on $\mathbf{W}$-each exposure $E_i$ becomes a deterministic (weighted) sum of random variables that do not correlate with unit $i$ 's potential outcomes. This construction satisfies both strong and weak unconfoundedness.

\subsection{The generalized propensity score}

We now introduce our proposed estimator based on the generalized propensity score (GPS), inspired by the literature on propensity scores for multivalued and continuous treatments \citep{hirano2004propensity}. All proofs of the lemmas and theorems in this section are provided in the appendix.

\begin{definition}[Generalized propensity score]\label{def:gps}
Let

$$
r(e, w)=f_{E \mid W}(e \mid w)
$$
be the conditional density of the exposure $E$ given the network $W$. We define the GPS as

$$
R=r(E, W)
$$
\end{definition}

Following the seminal work of \citet{rosenbaum1983central}, under weak unconfoundedness, conditioning on the GPS is sufficient to ensure the conditional independence of $\mathbb{I}\left\{E_i=e\right\}$ and $Y_i(e)$. We summarize this result in the two lemmas below.

\begin{lemma}[Balancing given the GPS]\label{lem:balancing}
For a specific exposure level $e$, if $r(e, W)$ is held constant, then the probability $\operatorname{Pr}(E=e)$ does not depend on the pre-exposure network $W$. Formally,

$$
W \perp \mathbb{I}(E=e) \mid r(e, W), \quad \forall e \in \mathcal{E}
$$
\end{lemma}

\begin{lemma}[Weak unconfoundedness given the GPS]\label{lem:unconf}
Under Assumption 2,

$$
\mathbb{I}(E=e) \perp Y(e) \mid r(e, W), \quad \forall e \in \mathcal{E}
$$
\end{lemma}

Lemma 2 primarily follows from Lemma 1 and is crucial for establishing the unbiasedness of our estimator. In essence, to achieve independence between the potential outcome $Y(e)$ and whether an outcome unit actually receives exposure $e$, it suffices to condition on the GPS at that exposure level. This strategy obviates the need to condition on the full vector $\mathbf{W}$, mitigating the risk of having sparse or empty strata when conditioning on a high-dimensional set of covariates.

We now introduce our principal theoretical result for unbiased estimation of the exposure-outcome function and its associated estimands.

\begin{theorem}[Modeling with the GPS]\label{thm:modeling}
Under previous assumptions, we have:

1. Outcome Model: Estimate the conditional expectation of the outcome as a function of the exposure level $e$ and the GPS $R$ :

$$
\beta(e, r)=\mathbb{E}[Y(e) \mid r(e, W)=r]=\mathbb{E}[Y(e) \mid E=e, R=r]
$$

2. Exposure-Outcome Function: Estimate $\mu(e)$ by taking the expectation of the above conditional expectation with respect to the GPS at exposure level $e$ :

$$
\mu(e)=\mathbb{E}[\beta(e, r(e, W))]
$$
\end{theorem}

Here, $\beta(e, r)$ does not have a direct causal interpretation; it is simply the conditional mean of the outcome given the exposure and the GPS. Some implementations of Theorem 1 require a parametric assumption on $\beta(e, r)$. However, as we show in the empirical section, semi-parametric approaches can also be employed.

Next, we present a Horvitz-Thompson-style result \citep{horvitz1952generalization}, which offers an alternative weighting approach for estimating the exposure-outcome function.

\begin{theorem}[Weighting by the GPS]\label{thm:weighting}
Under the preceding assumptions, one can estimate the exposure-outcome function $\mu(e)$ by averaging the reweighted responses at a particular exposure stratum $e$ :

$$
\mathbb{E}\left[\frac{Y \cdot \mathbb{I}(e)}{r(E, W)}\right]=\mathbb{E}[Y(e)]
$$
\end{theorem}

Unlike Theorem 1, Theorem 2 does not assume a specific parametric relationship between the response and the GPS. Nevertheless, Theorem 2 also lacks a strict causal interpretation. Both Theorem 1 and Theorem 2 leverage the GPS to estimate $\mu(e)$ at each exposure level $e$.

The function $r(E, W)$ is often ‘stabilized’ by the marginal probability $P_E(e)$ to prevent excessively large weights and thus control variance. However, such stabilization also presupposes positivity—namely, $\operatorname{Pr}(E=e)>0$ for the exposure levels of interest. Under the linear exposure assumption in a simple randomized design-where each $Z_j=1$ independently with probability $p_j \in(0,1)$-this condition $\operatorname{Pr}\left(E_i=e\right)>0$ typically holds only at the boundary exposures ( $e=0$ or $e=1$ ), and may fail for intermediate values of $e$.

Both Modeling with the GPS (Theorem 1) and Weighting by the GPS (Theorem 2) rely on the GPS to mitigate confounding bias. Specifically, Theorem 1 integrates the GPS into an outcome model, while Theorem 2 uses the GPS to weight observed outcomes. However, in practice, these approaches may still exhibit some residual bias related to confounders. Consequently, we introduce Theorem 3, which combines both modeling and weighting into a double deconfounding framework for further bias reduction and enhanced robustness.

\begin{lemma}[Weak unconfoundedness of the response model given the GPS]\label{lem:responsemodel}
Under the same assumptions, let $\beta(e, r)=\mathbb{E}[Y(e) \mid r(e, W)=r]$. Then,

$$
\mathbb{I}(E=e) \perp \beta(e, r) \mid r(e, W), \quad \forall e \in \mathcal{E}
$$
\end{lemma}

\begin{theorem}[Weighting and modeling with the GPS, or ``double deconfounded'']\label{thm:dd}
Under the previous assumptions, we can estimate $\mathbb{E}[Y(e)]$ by averaging the reweighted model-based predictions at exposure level $e$ :

$$
\mathbb{E}\left[\frac{\beta(e, r) \mathbb{I}(e)}{r(E, W)}\right]=\mathbb{E}[Y(e)]
$$
\end{theorem}

Here, $\beta(e, r)$ is a (possibly parametric) model of the outcome $Y ; \mathbb{I}(e)$ is an indicator function for exposure stratum $e$; and $r(E, W)$ is the GPS. Because Theorem 3 applies the GPS in both the modeling and weighting phases, we refer to this approach as "Double Deconfounding" (DD for short).

\subsection{Theoretical bounds on the proposed estimators}

\subsubsection{Lower bounds on the variance}

Under the linear exposure assumption, \citet{pouget2019variance} established that for any unbiased estimator $\hat{\tau}$ of the average treatment effect, its variance is bounded below by $\sigma^2/\Delta$, where $\Delta$ represents the expected empirical variance of the treatment exposure vector. Since our proposed estimators—GPS modeling, GPS weighting, and double deconfounding—are all unbiased under the stated assumptions, this theoretical lower bound applies to our setting. Moreover, they demonstrated that this bound can be approached through correlation clustering of diversion units, which maximizes the empirical variance $\Delta$. This suggests that combining our GPS-based estimation with an appropriate clustering scheme could potentially achieve near-optimal statistical efficiency in bipartite experiments.

\subsubsection{Upper bounds on the variance}

For an exposure level $e$, Theorem 2's estimator (the inverse-propensity-weighting estimator) is
$$
\hat{\mu}(e)=\frac{1}{n} \sum_{i=1}^n \frac{1\left(E_i=e\right) Y_i}{r\left(e, W_i\right)}
$$
where $1\left(E_i=e\right)$ is the indicator that unit $i$ is observed at exposure $e$, and $r\left(e, W_i\right)$ is the (generalized) propensity score: $r(e, w)=\operatorname{Pr}\left(E_i=e \mid W_i=w\right)$.

Under standard assumptions of bounded outcomes and positivity (detailed in Appendix), we can establish that
$$
\operatorname{Var}(\hat{\mu}(e)) \leq \frac{M^2}{\delta^2} \frac{\operatorname{Pr}(E=e)}{n}
$$
where $M$ is the bound on the outcome and $\delta$ is the lower bound on the propensity score. Since $\operatorname{Pr}(E=e) \leq 1$, this leads to a simpler (though slightly looser) bound:
$$
\operatorname{Var}(\hat{\mu}(e)) \leq \frac{M^2}{n \delta^2}
$$

Although the inverse-propensity-weighting estimator in Theorem 2 admits a relatively transparent upper bound on its variance, establishing analogous upper bounds for the regression-based estimator (Theorem 1) and the doubly deconfounded estimator (Theorem 3) is considerably more involved. In these latter estimators, modeling choices introduce additional sources of uncertainty, and the estimation errors from both the outcome model and the (generalized) propensity score must be jointly accounted for. As a result, practitioners often resort to alternative approaches, such as bootstrap methods, to obtain valid standard errors and confidence intervals, rather than attempting to derive a direct analytical upper bound. We will discuss these methods in detail in later sections.

\subsection{Considerations for observational data}

While we are primarily concerned with experimental settings, the results of this section are formulated in a way that makes them valid in observational settings as long as the unconfoundedness assumptions are satisfied. In practical terms, working with observational data usually implies two things: the functional form of the generalized propensity scores is unknown, so the generalized propensity scores must be estimated; and there is a variety of potential estimands of interest.

The first point is self-explanatory and expanded on in Section~7, but the second point merits discussion. In many experimental settings, the researchers are primarily interested in estimating $\mu(1)-\mu(0)$, the average effect of treating the whole population versus not treating anyone. When dealing with observational data, treating the whole population may not be feasible and the researchers might be interested in evaluating the cost-effectiveness of a marginal intervention, which is the case in, for example, \citet{papadogeorgou2019spatial}.

\section{Estimation algorithms}
In Section 3, we established three theoretical approaches for unbiased causal effect estimation in bipartite experiments: modeling with the GPS (Theorem 1), weighting by the GPS (Theorem 2), and double deconfounding (Theorem 3). This section presents practical algorithms for implementing these approaches. We first discuss algorithms for static networks, followed by necessary modifications for dynamic network settings.

\subsection{Algorithms for static networks}

Our goal is to estimate the marginal exposure-outcome effect, denoted as $\tau$, which represents the average change in outcome per unit change in exposure. Since we treat exposure $E_i$ as a continuous variable, this effect can be estimated through the model parameter $\theta_1$ corresponding to the exposure term. Table 4.1 summarizes our proposed algorithms, their theoretical foundations, and their key characteristics.

\begin{table}[htbp]
\small
\centering
\caption{Summary of Estimation Algorithms}
\begin{tabular}{lll}
\hline
\textbf{Algorithm} & \textbf{Theoretical Basis} & \textbf{Description} \\ \hline
Naïve Regression & None (Baseline) & Simple linear regression \\
Two-Stage Least Squares (TSLS) & Theorem 1 & GPS-adjusted instrumental variables \\
Model Regression & Theorem 1 & Direct GPS conditioning \\
Generalized Additive Model (GAM) & Theorem 1 & Flexible functional forms \\
Inverse Probability Weighting (IPW) & Theorem 2 & Classic propensity weighting \\
IPW Boosting & Theorem 2 & Machine learning enhanced weighting \\
Double Deconfounded TSLS & Theorem 3 & Combined modeling and weighting with TSLS \\
Double Deconfounded GAM & Theorem 3 & Combined modeling and weighting with GAM \\ \hline
\end{tabular}
\label{tab:estimation-algorithms}
\end{table}

\subsubsection{Naïve Regression (Baseline)}

We begin with a simple baseline approach that ignores potential network-induced confounding. This approach assumes a linear relationship between exposure and outcome through the following normal model:

\begin{equation}
Y_i \mid E_i \sim \mathcal{N}(\theta_0 + \theta_1 E_i, \sigma_y^2)
\end{equation}

The marginal exposure-outcome effect $\tau$ is estimated by the ordinary least squares estimator for $\theta_1$:

\begin{equation}
\hat{\tau} = \hat{\theta}_1 = (E'E)^{-1}E'Y = \frac{\sum_{i=1}^N(E_i - \bar{E})(Y_i - \bar{Y})}{\sum_{i=1}^N(E_i - \bar{E})^2}
\end{equation}

While this estimator provides a useful benchmark, it typically yields biased estimates due to its failure to account for network-induced confounding. The bias arises because the exposure $E_i$ is not randomly assigned but rather determined by the network structure and treatment assignment mechanism.

\subsubsection{Two Stage Least Square (TSLS)}

Building on Theorem 1 (Model with the GPS), we implement a two-stage least squares approach that uses the GPS as an instrumental variable. This approach proceeds in two stages:

First Stage: Exposure Model

We begin by modeling the exposure as a function of network characteristics:
\begin{equation}
E_i = \alpha_0 + \alpha_1|W_i| + \epsilon_i
\end{equation}
where $|W_i|$ represents the network degree of unit $i$. Given the continuous nature of exposure, we employ a Gaussian regression model rather than logistic regression.

From this model, we compute the residuals:
\begin{equation}
\hat{\gamma}_i = E_i - \hat{\alpha}_0 - \hat{\alpha}_1|W_i|
\end{equation}

The GPS is then estimated using these residuals:
\begin{equation}
\hat{R}_i(\hat{\gamma}_i) = \phi(\hat{\gamma}_i) = \frac{1}{\sqrt{2\pi\hat{\sigma}^2}} \exp\left[-\frac{1}{2\hat{\sigma}^2}\hat{\gamma}_i^2\right]
\end{equation}
where $\phi$ denotes the Gaussian probability density function.

Second Stage: Outcome Model

We specify the TSLS model as:
\begin{equation}
\beta_\theta(E_i, R_i(\gamma_i)) = \theta_0 + \theta_1E_i + \theta_2(E_i - \gamma_i)
\end{equation}

The expected potential outcome for exposure level $e$ is then estimated as:
\begin{equation}
\hat{\mathbb{E}}[Y(e)] = \frac{1}{N_e}\sum_{i=1}^{N_e}[\hat{\theta}_0 + \hat{\theta}_1E_i + \hat{\theta}_2(E_i - \hat{\gamma}_i)] \cdot \mathbb{I}(E_i = e)
\end{equation}
where $N_e = \sum_{i=1}^N \mathbb{I}(E_i = e)$ is the number of units with exposure level $e$.

The marginal exposure-outcome effect is estimated by $\hat{\tau} = \hat{\theta}_1$, representing the average change in outcome per unit change in exposure after accounting for network confounding through the GPS.

\subsubsection{Model Regression}

Building on the GPS estimation previously, we implement a direct conditioning approach that explicitly models the outcome as a function of both exposure and the GPS.

Using the residuals from the exposure model, we estimate the GPS as:
\begin{equation}
\hat{R}_i(\hat{\gamma}_i) = \frac{1}{\sqrt{2\pi\hat{\sigma}^2}} \exp\left[-\frac{1}{2\hat{\sigma}^2}\hat{\gamma}_i^2\right]
\end{equation}

We specify a parametric model for the conditional expectation of the outcome given exposure and GPS:
\begin{equation}
\mathbb{E}[Y_i \mid E_i, R_i(\gamma_i)] = \beta_\theta(E_i, R_i(\gamma_i))
\end{equation}

Specifically, we employ a quadratic specification that allows for both main effects and interactions:
\begin{equation}
\beta_\theta(E_i, R_i(\gamma_i)) = \theta_0 + \theta_1E_i + \theta_2R_i + \theta_3E_i^2 + \theta_4R_i^2 + \theta_5E_iR_i
\end{equation}

The average potential outcome at exposure level $e$ is then estimated as:
\begin{equation}
\hat{\mathbb{E}}[Y(e)] = \frac{1}{N_e}\sum_{i=1}^{N_e}[\hat{\beta}_\theta(E_i, R_i(\gamma_i))] \cdot \mathbb{I}(E_i = e)
\end{equation}
where $N_e = \sum_{i=1}^N \mathbb{I}(E_i = e)$.

The marginal exposure-outcome effect is estimated by $\hat{\tau} = \hat{\theta}_1$, representing the linear component of the exposure effect after controlling for the GPS.

The effectiveness of this approach relies critically on the correct specification of $\beta_\theta(E_i, R_i(\gamma_i))$. While our quadratic specification allows for some nonlinearity and interaction effects, it remains a parametric approximation. Alternative specifications are possible within this framework, but all purely model-based approaches face potential bias from model misspecification. We examine these concerns empirically in Section~6.

\subsubsection{Generalized Additive Models (GAM)}

Following the GPS calculation procedure outlined in the previous algorithms, we introduce a more flexible approach using generalized additive models. This method combines parametric and non-parametric components to capture complex relationships between the outcome, exposure, and GPS.

We specify a semi-parametric model that maintains a linear term for the exposure while allowing flexible functional forms for other components:
\begin{equation}
\beta_\theta(E_i, R_i(\gamma_i)) = \theta_1E_i + s(R_i) + s(E_i^2) + s(R_i^2) + s(E_iR_i)
\end{equation}
where $s(\cdot)$ denotes a smooth function implemented using splines. This specification retains a parametric coefficient $\theta_1$ for the main exposure effect, while using smooth functions to capture potential nonlinear relationships in the data. The inclusion of smooth functions for squared terms and interactions allows flexible modeling of complex relationships between exposure and GPS.

The marginal exposure-outcome effect is estimated by $\hat{\tau} = \hat{\theta}_1$, representing the linear component of the exposure effect. This parameter has a clear causal interpretation while the model maintains flexibility through its non-parametric components.

For applications focused solely on estimating the average exposure-outcome function $\mu(e) = \mathbb{E}[Y(e)]$ at specific exposure levels, researchers may consider a fully non-parametric specification by replacing the linear exposure term with a smooth function $s(E_i)$. This approach sacrifices the direct interpretability of a marginal effect estimate but may capture more complex exposure-outcome relationships.

\subsubsection{Inverse Probability Weighting (IPW)}

Building on Theorem 2 (Weighting with the GPS), we implement an inverse probability weighting approach that proceeds in three stages: GPS estimation, weight stabilization, and weighted regression.

We begin with the exposure model:
\begin{equation}
E_i = \alpha_0 + \alpha_1|W_i| + \epsilon_i
\end{equation}

From this model, we compute the residuals:
\begin{equation}
\hat{\gamma}_i = E_i - \hat{\alpha}_0 - \hat{\alpha}_1|W_i|
\end{equation}

The GPS is then estimated using a Gaussian kernel:
\begin{equation}
\hat{R}_i = \frac{1}{\sqrt{2\hat{\sigma}^2}} \exp\left[-\frac{1}{2\hat{\sigma}^2}\hat{\gamma}_i^2\right]
\end{equation}

To improve efficiency, we stabilize the weights using the marginal probability of exposure:
\begin{equation}
R'_i = \frac{\hat{P}_E(E_i)}{\hat{R}_i}
\end{equation}
where $\hat{P}_E(E_i)$ represents the estimated marginal probability of receiving exposure level $E_i$.

Following the Marginal Structure Model (MSM) from \citet{robins2000marginal}, we estimate the marginal exposure-outcome effect using a weighted regression approach. For a given exposure level $e$, the average potential outcome is estimated as:
\begin{equation}
\hat{\mathbb{E}}[Y(e)] = \frac{1}{N_e}\sum_{i=1}^{N_e}\frac{Y_i \cdot \mathbb{I}(E_i = e)}{R'_i}
\end{equation}

The marginal exposure-outcome effect is then estimated by minimizing the weighted least squares objective:
\begin{equation}
\sum_{i=1}^N R'_i(Y_i - \theta_0 - \theta_1E_i)^2
\end{equation}

The resulting coefficient $\hat{\tau} = \hat{\theta}_1$ represents the average causal effect of a unit change in exposure. This approach provides a non-parametric alternative to the model-based methods, relying instead on the weighting principle established in Theorem 2 to achieve unbiased estimation.

\subsubsection{IPW with Gradient Boosted Regression Trees (IPW Boost)}

Building upon the IPW framework from the previous section, we enhance the exposure modeling stage using gradient boosted regression trees while maintaining the same weighting principles from Theorem 2.

Instead of the linear exposure model, we employ a gradient boosted regression tree model:
\begin{equation}
E_i = \sum_{tr=1}^{Tr} \delta_{tr} \times f_{tr}(W_{i.})
\end{equation}
where $Tr$ denotes the number of boosting rounds.
    $f_{tr}(\cdot)$ represents the prediction from the $tr$-th base tree model,
     $\delta_{tr}$ is the weight assigned to the $tr$-th tree,
     $W_{i.}$ represents the network characteristics for unit $i$

Using the boosted model's predictions, we compute the residuals:
\begin{equation}
\hat{\gamma}_i = E_i - \hat{E}_i
\end{equation}

The GPS is then estimated using a Gaussian kernel:
\begin{equation}
\hat{R}_i = \frac{1}{\sqrt{2\pi\hat{\sigma}^2}} \exp\left[-\frac{1}{2\hat{\sigma}^2}\hat{\gamma}_i^2\right]
\end{equation}

As in the standard IPW approach, we stabilize the weights:
\begin{equation}
R'_i = \frac{\hat{P}_E(E_i)}{\hat{R}_i}
\end{equation}

Following the same approach as standard IPW, we estimate the average potential outcome for exposure level $e$:
\begin{equation}
\hat{\mathbb{E}}[Y(e)] = \frac{1}{N_e}\sum_{i=1}^{N_e}\frac{Y_i \cdot \mathbb{I}(E_i = e)}{R'_i}
\end{equation}

The marginal exposure-outcome effect is estimated by minimizing:
\begin{equation}
\sum_{i=1}^N R'_i(Y_i - \theta_0 - \theta_1E_i)^2
\end{equation}

The resulting coefficient $\hat{\tau} = \hat{\theta}_1$ represents the causal effect of a unit change in exposure. This boosted approach offers increased flexibility in modeling the exposure mechanism while maintaining the non-parametric weighting framework for outcome estimation.

\subsubsection{Double Deconfounded Two-Stage Least Squares (DD TSLS)}

Building on Theorem 3 (Model and Weight with the GPS), we combine the flexible GPS estimation from IPW Boost with the instrumental variable approach from TSLS. This double deconfounding strategy leverages both modeling and weighting to enhance robustness against confounding bias.

We utilize the GPS ($R_i$) and stabilized weights ($R'_i$) as computed in the IPW Boost algorithm, taking advantage of the machine learning approach to exposure modeling.

We specify a weighted outcome model that incorporates both exposure and residuals:
\begin{equation}
\mathbb{E}[Y_i \mid E_i, R_i(\gamma_i)] = \beta_\theta(E_i, R_i(\gamma_i)) = \theta_0 + \theta_1E_i + \theta_2(E_i - \hat{\gamma}_i)
\end{equation}
where $\hat{\gamma}_i$ represents the residuals from the boosted exposure model.

For a given exposure level $e$, we estimate the average potential outcome by combining the model predictions with inverse probability weighting:
\begin{equation}
\hat{\mathbb{E}}[Y(e)] = \frac{1}{N_e}\sum_{i=1}^{N_e}\left(\frac{\hat{\theta}_0 + \hat{\theta}_1E_i + \hat{\theta}_2(E_i - \hat{\gamma}_i)}{R'_i}\right) \cdot \mathbb{I}(E_i = e)
\end{equation}

The marginal exposure-outcome effect is estimated by $\hat{\tau} = \hat{\theta}_1$. This approach benefits from double robustness: it remains consistent if either the exposure model (through the GPS) or the outcome model is correctly specified.

\subsubsection{Double Deconfounded Generalized Additive Model (DD GAM)}

Building on Theorem 3 (Model and Weight with the GPS), we combine the machine learning-based GPS estimation from IPW Boost with the flexible modeling approach of GAM. This synthesis provides a highly adaptive framework for causal effect estimation.

We utilize the GPS ($R_i$) and stabilized weights ($R'_i$) as computed in the IPW Boost algorithm, leveraging gradient boosted trees for exposure modeling.

We specify a weighted semi-parametric model that maintains a linear term for the primary exposure effect while allowing flexible functional forms for other components:
\begin{equation}
\beta_\theta(E_i, R_i(\gamma_i)) = \theta_1E_i + s(R_i) + s(E_i^2) + s(R_i^2) + s(E_iR_i)
\end{equation}
where $s(\cdot)$ denotes smooth functions implemented through splines. This specification combines the interpretability of a linear exposure effect with the flexibility of non-parametric components for complex relationships and interactions.

For a given exposure level $e$, we estimate the average potential outcome by combining the GAM predictions with inverse probability weighting:
\begin{equation}
\hat{\mathbb{E}}[Y(e)] = \frac{1}{N_e}\sum_{i=1}^{N_e}\left(\frac{\theta_1E_i + s(R_i) + s(E_i^2) + s(R_i^2) + s(E_iR_i)}{R'_i}\right) \cdot \mathbb{I}(E_i = e)
\end{equation}

The marginal exposure-outcome effect is estimated by $\hat{\tau} = \hat{\theta}_1$. This approach combines the advantages of non-parametric flexibility in both the exposure modeling (through boosting) and outcome modeling (through GAM) stages while maintaining the double robustness property of Theorem 3.

\subsection{Algorithms for dynamic networks}

In dynamic network settings, the exposure-outcome relationship may evolve over time due to factors such as learning effects, network formation processes, and temporal dependencies. To account for these dynamics, we modify our estimation approaches to incorporate time-varying effects while maintaining the core principles established in our theoretical framework.

\subsubsection{Naïve Regression (Baseline)}

For dynamic networks, we modify the baseline naïve regression to incorporate time-varying effects. The conditional distribution of the outcome $Y_{it}$ given exposure $E_{it}$ is assumed to follow a normal distribution:

\begin{equation}
Y_{it} \mid E_{it} \sim \mathcal{N}\left(\theta_0 + \theta_1E_{it} + \sum_{t \in T} \xi_t FT_{it}, \sigma_y^2\right)
\end{equation}

where  $\theta_0$ represents the global intercept,
     $\theta_1$ captures the exposure effect,
     $\xi_t$ denotes the time-specific effect for period $t$,
    $FT_{it}$ is an indicator variable for unit $i$ being observed in time period $t$,
    $\sigma_y^2$ represents the variance of the outcome.

The introduction of dynamic intercepts $\sum_{t \in T} \xi_t FT_{it}$ allows the model to capture temporal patterns without imposing restrictive parametric assumptions about the form of time dependence. As shown in Figure 3, these dynamic intercepts reveal an S-shaped learning curve pattern that can be characterized by the logistic function:
$\frac{\kappa}{1 + e^{-\lambda(\tau - \tau_0)}}$,
where $\kappa$ represents the carrying capacity (maximum achievable level),
    $\lambda$ denotes the growth rate,
     $\tau_0$ indicates the time of median growth.

This emergent S-shaped pattern suggests a natural learning process in the network, characterized by slow initial growth, followed by rapid acceleration, and eventual saturation. However, by using time-specific fixed effects rather than imposing this functional form directly, our model maintains flexibility while still capturing the underlying temporal dynamics.

\begin{figure}[htbp]
    \centering
    \includegraphics[width=0.8\textwidth]{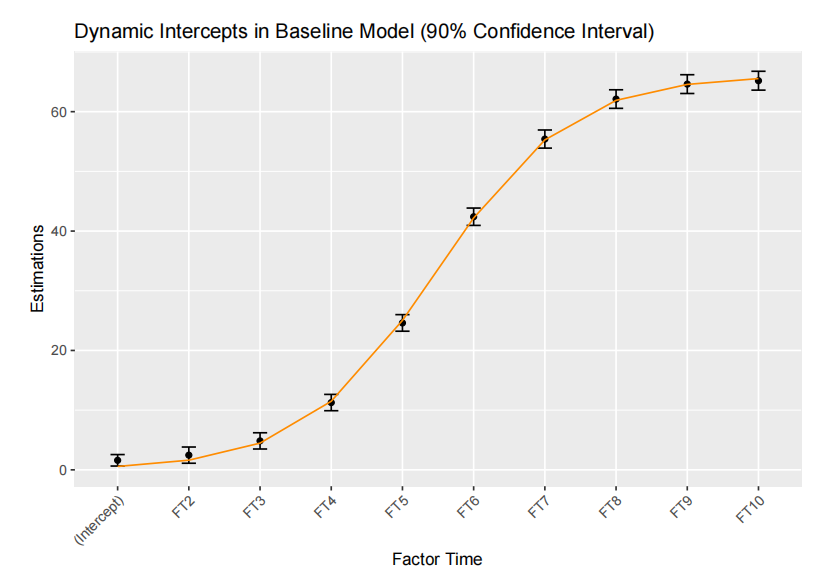}
    \caption{Dynamic Networks Simulation: Dynamic Intercepts of Baseline Algorithm}
    \label{fig:learning-curve}
\end{figure}

The marginal exposure-outcome effect in this dynamic setting is still estimated by $\hat{\tau} = \hat{\theta}_1$, but now represents the average treatment effect after accounting for temporal patterns in the data. While this naïve approach provides a useful baseline for comparison, it may still yield biased estimates due to its failure to account for network-induced confounding beyond simple time effects.

\subsubsection{Two Stage Least Square (TSLS)}

The Two Stage Least Square (TSLS) algorithm maintains its fundamental structure when applied to dynamic networks, preserving the robustness properties established in the static network setting. In this context, the estimation process proceeds as follows:

\begin{equation}
\begin{aligned}
\text{First Stage: } & E_{it} = \alpha_0 + \alpha_1|W_{it}| + \gamma_{it} \\
\text{Second Stage: } & Y_{it} = \theta_0 + \theta_1E_{it} + \theta_2(E_{it} - \gamma_{it}) + \epsilon_{it}
\end{aligned}
\end{equation}

A key feature of TSLS in dynamic networks is that the temporal dependencies are naturally captured through the residuals $\gamma_{it}$ from the exposure model. This characteristic eliminates the need for explicit dynamic intercept terms of the form $\sum_{t \in T} \xi_t FT_{it}$ that were necessary in the naïve regression approach.
The rationale behind this property stems from the instrumental variables framework underlying TSLS. The residuals $\gamma_{it}$ serve as a control function that accounts for both network-induced confounding and temporal evolution of the exposure-outcome relationship. Consequently, the exposure effect estimator $\hat{\theta}_1$ maintains its interpretation as the causal effect of interest, now incorporating both cross-sectional and temporal variation in the data.

This inherent robustness to temporal dynamics represents a significant advantage of TSLS-based approaches in dynamic network settings, as it reduces the risk of misspecification in the temporal component of the model while still accounting for time-varying confounding effects.

\subsubsection{Model Regression}

For dynamic networks, we extend the static network GPS methodology while incorporating temporal effects. The procedure follows several steps:

First, we specify the exposure model:
\begin{equation}
E_i = \alpha_0 + \alpha_1|W_i| + \epsilon_i
\end{equation}

Given the continuous nature of exposure $E_i$, we employ a Gaussian regression model rather than logistic regression, though alternative Generalized Linear Models (GLM) such as multinomial regression could be considered. 

From this model, we compute the residuals:
\begin{equation}
\hat{\gamma}_i = E_i - \hat{\alpha}_0 - \hat{\alpha}_1|W_i|
\end{equation}

The GPS is then estimated using a Gaussian kernel:
\begin{equation}
R_i(\hat{\gamma}_i) = \phi(\hat{\gamma}_i) = \frac{1}{\sqrt{2\pi\hat{\sigma}^2}} \exp\left[-\frac{1}{2\hat{\sigma}^2}\hat{\gamma}_i^2\right]
\end{equation}
where $\phi$ denotes the probability density function of the normal distribution.

For the outcome model, we modify the regression equation to include dynamic intercepts:
\begin{equation}
\begin{split}
\beta_\theta(E_i, R_i(\gamma_i)) = \theta_0 + \theta_1E_i + \theta_2R_i + \theta_3E_i^2 + \theta_4R_i^2 + \theta_5E_iR_i + \sum_{t \in T} \xi_tFT_{it}
\end{split}
\end{equation}
where $\sum_{t \in T} \xi_tFT_{it}$ captures the time-varying effects.

The average potential outcome for exposure level $e$ is then estimated as:
\begin{equation}
\begin{split}
\mathbb{E}[Y(e)] = \frac{1}{N_e}\sum_{i=1}^{N_e}[\theta_0 + \theta_1E_i + \theta_2R_i + \theta_3E_i^2 + \theta_4R_i^2 + \theta_5E_iR_i + \sum_{t \in T} \xi_tFT_{it}] \cdot \mathbb{I}(E_i = e)
\end{split}
\end{equation}

Finally, we estimate the marginal exposure-outcome effect as $\hat{\tau} = \hat{\theta}_1$. This coefficient represents the linear component of the exposure effect after accounting for both the GPS adjustment and temporal dynamics in the network structure.

\subsubsection{Generalized Additive Models (GAM)}

Following the static network methodology, we employ the Generalized Propensity Score (GPS) within a semi-parametric framework. The procedure begins with the exposure model:

\begin{equation}
E_i = \alpha_0 + \alpha_1|W_i| + \epsilon_i
\end{equation}

From this model, we compute the residuals:
\begin{equation}
\hat{\gamma}_i = E_i - \hat{\alpha}_0 - \hat{\alpha}_1|W_i|
\end{equation}

The GPS is then estimated using:
\begin{equation}
R_i(\hat{\gamma}_i) = \phi(\hat{\gamma}_i) = \frac{1}{\sqrt{2\pi\hat{\sigma}^2}} \exp\left[-\frac{\hat{\gamma}_i^2}{2\hat{\sigma}^2}\right]
\end{equation}

For dynamic networks, we modify the outcome model to incorporate both smooth functions and temporal effects:
\begin{equation}
\begin{split}
\beta_\theta(E_i, R_i(\gamma_i)) = \theta_1E_i + s(R_i) + s(E_i^2) + s(R_i^2) + s(E_iR_i) + \sum_{t \in T} \xi_tFT_{it}
\end{split}
\end{equation}
where $s(\cdot)$ denotes smooth functions implemented through splines, and $\sum_{t \in T} \xi_tFT_{it}$ captures the dynamic intercepts.

The average potential outcome for exposure level $e$ is estimated as:
\begin{equation}
\begin{split}
\mathbb{E}[Y(e)] = \frac{1}{N_e}\sum_{i=1}^{N_e}[\theta_1E_i + s(R_i) + s(E_i^2) + s(R_i^2) + s(E_iR_i) + \sum_{t \in T} \xi_tFT_{it}] \cdot \mathbb{I}(E_i = e)
\end{split}
\end{equation}

The marginal exposure-outcome effect is estimated as $\hat{\tau} = \hat{\theta}_1$, representing the linear component of the exposure effect while allowing for flexible functional forms in other components of the model.

Notably, in the dynamic network context, the incorporation of time-varying effects through $\sum_{t \in T} \xi_tFT_{it}$ necessitates a modification of our estimation strategy. Specifically, algorithms originally based on Theorem 2 (Weighting by the GPS) are adapted to follow Theorem 3 (Weighting and Modeling with the GPS) to properly account for these dynamic intercepts. This adaptation ensures that both the temporal evolution of the network and the flexible functional relationships between variables are appropriately captured in the estimation process.

\subsubsection{Inverse Probability Weighting (IPW)}

For the dynamic network setting, we implement IPW using the Generalized Propensity Score (GPS). The estimation proceeds in several stages:

First, we specify the exposure model:
\begin{equation}
E_i = \alpha_0 + \alpha_1|W_i| + \epsilon_i
\end{equation}

From this model, we compute the residuals:
\begin{equation}
\hat{\gamma}_i = E_i - \hat{\alpha}_0 - \hat{\alpha}_1|W_i|
\end{equation}

The GPS is then estimated using:
\begin{equation}
R_i(\hat{\gamma}_i) = \phi(\hat{\gamma}_i) = \frac{1}{\sqrt{2\pi\hat{\sigma}^2}} \exp\left[-\frac{\hat{\gamma}_i^2}{2\hat{\sigma}^2}\right]
\end{equation}

We stabilize the weights using:
\begin{equation}
R'_i = \frac{\hat{P}_E(E_i)}{\hat{R}_i}
\end{equation}

For each time period in the factor set $FT_t \in (t \in T)$, we estimate dynamic intercept terms $\xi_t$. Assuming a parametric relationship between the response $Y_{it}$ and Factor Time $FT_{it}$, we define the conditional expectation $\mathbb{E}[Y_{it} \mid FT_{it}]$ as:
\begin{equation}
\beta_\xi(FT_{it}) = \sum_{t \in T} \xi_tFT_{it}
\end{equation}

After estimating the parameters $\hat{\xi}_t$, we compute the average potential outcome for exposure level $e$:
\begin{equation}
\mathbb{E}[Y(e)] = \frac{1}{N_e}\sum_{i=1}^{N_e} \frac{\beta_\xi(FT_{it}) \cdot \mathbb{I}(E_{it} = e)}{R'_{it}}
\end{equation}

Following Theorem 2, we employ a Marginal Structural Model (MSM) to estimate the marginal exposure-outcome effect. The MSM fits a weighted linear regression of outcomes $(Y_i)$ on exposure $(E_i)$, using the stabilized GPS $(R'_i)$ as weights. The objective function to be minimized is:
\begin{equation}
\sum_{i=1}^N R'_i(Y_i - \sum_{t \in T} \xi_tFT_{it} - \theta_0 - \theta_1E_i)^2
\end{equation}

The marginal exposure-outcome effect is then estimated as $\hat{\tau} = \hat{\theta}_1$. This estimator accounts for both the temporal evolution of the network through the dynamic intercepts and the confounding effects through the GPS weights.

\subsubsection{IPW with Gradient Boosted Regression Tree Model (IPW Boost)}

We modify the standard IPW approach by replacing the Gaussian linear exposure model with a Gradient Boosted Regression Tree Model. In this framework, the exposure $E_i$ is modeled as:

\begin{equation}
E_i = \sum_{tr=1}^{Tr} \delta_{tr} \times f_{tr}(W_{i})
\end{equation}

where $Tr$ denotes the number of boosting rounds,
     $f_{tr}(\cdot)$ represents the prediction from the $tr$-th base tree model,
    $\delta_{tr}$ is the weight assigned to the $tr$-th tree,
    $W_{i}$ represents the network characteristics for unit $i$.

Using the boosted model's predictions $\hat{E}_i$, we compute the residuals:
\begin{equation}
\hat{\gamma}_i = E_i - \hat{E}_i
\end{equation}

The GPS is then estimated using:
\begin{equation}
R_i(\hat{\gamma}_i) = \phi(\hat{\gamma}_i) = \frac{1}{\sqrt{2\pi\hat{\sigma}^2}} \exp\left[-\frac{\hat{\gamma}_i^2}{2\hat{\sigma}^2}\right]
\end{equation}

As in the standard IPW approach, we stabilize the weights:
\begin{equation}
R'_i = \frac{\hat{P}_E(E_i)}{\hat{R}_i}
\end{equation}

The potential outcome model maintains the same structure as in the standard IPW:
\begin{equation}
\beta_\xi(FT_{it}) = \sum_{t \in T} \xi_tFT_{it}
\end{equation}

Similarly, the computation of average potential outcomes for exposure level $e$ follows the standard IPW methodology. The marginal exposure-outcome effect is estimated using the Marginal Structural Model (MSM), yielding $\hat{\tau} = \hat{\theta}_1$. 

This boosted approach offers increased flexibility in modeling the exposure mechanism while maintaining the established weighting framework for outcome estimation. The use of gradient boosted trees allows for automatic capture of non-linear relationships and interactions in the exposure model, potentially improving the accuracy of the GPS estimates.

\subsubsection{IPW Boost and TSLS (DD TSLS)}

For dynamic networks, the DD TSLS algorithm maintains its structure from the static network case, as the exposure model residuals inherently capture temporal dynamics. Consequently, there is no need to explicitly include dynamic intercepts $\sum_{t \in T} \xi_tFT_{it}$ in the outcome model.

We begin with a Gradient Boosted Regression Tree Model for the exposure:
\begin{equation}
E_i = \sum_{tr=1}^{Tr} \delta_{tr} \times f_{tr}(W_{i})
\end{equation}

where $Tr$ represents the number of boosting rounds,
    $f_{tr}(\cdot)$ is the prediction from the $tr$-th base tree,
     $\delta_{tr}$ denotes the weight for the $tr$-th tree,
    $W_{i}$ captures the network characteristics.

Using the model's predictions $\hat{E}_i$, we compute residuals $\hat{\gamma}_i$ and estimate the GPS:
\begin{equation}
R_i(\hat{\gamma}_i) = \phi(\hat{\gamma}_i) = \frac{1}{\sqrt{2\pi\hat{\sigma}^2}} \exp\left[-\frac{\hat{\gamma}_i^2}{2\hat{\sigma}^2}\right]
\end{equation}

We then stabilize the weights:
\begin{equation}
R'_i = \frac{\hat{P}_E(E_i)}{\hat{R}_i}
\end{equation}

The outcome model, conditional on exposure and residuals $(\hat{\gamma}_i)$, is specified as:
\begin{equation}
\mathbb{E}[Y_i \mid E_i, R_i(\gamma_i)] = \beta_\theta(E_i, R_i(\gamma_i)) = \theta_0 + \theta_1E_i + \theta_2(E_i - \hat{\gamma}_i)
\end{equation}

After estimating the model parameters $\hat{\theta}$, we compute the average potential outcome for exposure level $e$:
\begin{equation}
\mathbb{E}[Y(e)] = \frac{1}{N_e}\sum_{i=1}^{N_e}\left(\frac{\hat{\theta}_0 + \hat{\theta}_1E_i + \hat{\theta}_2(E_i - \hat{\gamma}_i)}{R'_i}\right) \cdot \mathbb{I}(E_i = e)
\end{equation}

The marginal exposure-outcome effect is estimated as $\hat{\tau} = \hat{\theta}_1$. This combined approach leverages both the flexibility of gradient boosting for exposure modeling and the robustness of TSLS for causal effect estimation, while inherently accounting for temporal dynamics through the residuals.

\subsubsection{IPW Boost and GAM (DD GAM)}

Following the structure of the DD TSLS algorithm, we incorporate a semiparametric Generalized Additive Model (GAM) for the outcome model in the final stage.

We begin with a Gradient Boosted Regression Tree Model for the exposure:
\begin{equation}
E_i = \sum_{tr=1}^{Tr} \delta_{tr} \times f_{tr}(W_{i.})
\end{equation}

where $f_{tr}(\cdot)$ represents the prediction from the $tr$-th base tree model and $\delta_{tr}$ denotes its corresponding weight.

From the model's predictions $\hat{E}_i$, we compute the residuals and estimate the GPS:
\begin{equation}
R_i(\hat{\gamma}_i) = \phi(\hat{\gamma}_i) = \frac{1}{\sqrt{2\pi\hat{\sigma}^2}} \exp\left[-\frac{\hat{\gamma}_i^2}{2\hat{\sigma}^2}\right]
\end{equation}

We then stabilize the weights:
\begin{equation}
R'_i = \frac{\hat{P}_E(E_i)}{\hat{R}_i}
\end{equation}

The outcome model is specified as a semiparametric function:
\begin{equation}
\begin{split}
\beta_\theta(E_i, R_i(\gamma_i)) = \theta_1E_i + s(R_i) + s(E_i^2) + s(R_i^2) + s(E_iR_i) + \sum_{t \in T} \xi_tFT_{it}
\end{split}
\end{equation}
where $s(\cdot)$ denotes smooth functions implemented through splines.

After estimating the model parameters $\hat{\theta}$, we compute the average potential outcome for exposure level $e$:
\begin{equation}
\begin{split}
\mathbb{E}[Y(e)] = \frac{1}{N_e}\sum_{i=1}^{N_e}\left(\frac{\theta_1E_i + s(R_i) + s(E_i^2) + s(R_i^2) + s(E_iR_i) + \sum_{t \in T} \xi_tFT_{it}}{R'_i}\right) \cdot \mathbb{I}(E_i = e)
\end{split}
\end{equation}

The marginal exposure-outcome effect is estimated as $\hat{\tau} = \hat{\theta}_1$.

The application of dynamic networks proves particularly valuable in understanding customer behavior patterns, especially in contexts involving the dissemination of new product information, advertisements, or pricing strategies. This framework explicitly acknowledges the temporal dimension of customer learning and adaptation. When new information is introduced into a market—whether through product launches, promotional campaigns, or price adjustments—there exists a natural lag between information introduction and customer assimilation.
This delay represents a critical learning period during which customers
process and internalize new information, 
    develop updated preferences and demands,
    and form new network connections based on these evolved preferences.
The dynamic network framework captures this evolutionary process, modeling how customer interactions and preferences adapt over time. This temporal aspect is particularly crucial in scenarios where network formation itself is endogenous to the learning process, with new connections emerging as customers discover and respond to new information in the marketplace.

\section{Variance estimation}

The proposed approach can be considered practical only if it provides a way to estimate confidence intervals for the parameter of interest. One simple way to estimate variance is to treat the model as a simple regression problem, ignoring the dependence of exposures across outcome units. For example, the ``na\"ive bootstrap'' method would sample individual observations $(Y_i, E_i, W_i)$ with replacement, computing for each sample set a value for the estimator and constructing the confidence interval using the quantiles of the resulting distribution.

We begin by showing that these standard variance estimators lead to correct coverage probabilities under a general potential outcomes model with i.i.d.\ error terms:
\begin{equation}\label{eq:uncorrmodel}
Y = \Phi(W, E)\beta + \varepsilon,
\end{equation}
where $\Phi$ is a parametric function of the graph weights $W_i$ and exposure $E_i$, subject to certain regularity conditions, $\beta$ and $\varepsilon$ are vectors of dimensions $K$ and $N$ respectively, and the error term $\varepsilon$ verifies $\mathbb{E}[\varepsilon \mid \Phi(W,E)] = 0$ and $\Var(\varepsilon \mid \Phi(W,E)) = \sigma^2_\varepsilon I_N$, where $I_N$ is the $N \times N$ identity matrix. We refer the reader to the Appendix for the proof of the following theorem and a discussion of the regularity conditions on $\Phi(W_i, E_i)$.

\begin{theorem}[Na\"ive bootstrap validity under uncorrelated errors]\label{thm:naiveboot}
Under the response model of Equation~\ref{eq:uncorrmodel}, both the na\"ive bootstrap- and the asymptotic OLS-based methods lead to valid confidence intervals.
\end{theorem}

The assumption of uncorrelated error terms may not be tenable in many cases. In the context of market experiments discussed in the introduction, a seller may change the price of an item, affecting the total amount $Y_i$ spent by every buyer $i$ that buys from that seller. To capture these correlated error terms, we consider a more general model:
\begin{equation}\label{eq:corrmodel}
Y = \Phi(W, E)\beta + W\gamma + \varepsilon,
\end{equation}
where the correlation is introduced through the additional $W\gamma$ term. Let $u = W\gamma + \varepsilon$, such that the response model can be more concisely written as $Y = \Phi(W, E)\beta + u$. To avoid identification issues for $\beta$, we impose that $\gamma = \{\gamma_j\}$ are i.i.d.\ normal, with mean 0 and variance $\sigma^2_\gamma$.

\begin{theorem}[Failure of the na\"ive bootstrap under correlated errors]\label{thm:naivefail}
Under some regularity assumptions on $\Phi(W_i, E_i)$ (discussed in the Appendix) and the response model of Equation~\ref{eq:corrmodel}, we have:
$$
\Var\left(\sqrt{N}(\hat\beta - \beta) \,\middle|\, E, W\right) = \sigma^2_\varepsilon Q_\Phi^{-1} + \sigma^2_\gamma Q_\Phi^{-1} Q_{\Phi W} Q_\Phi^{-1},
$$
where $Q_\Phi = N^{-1}\Phi^{T}\Phi$ and $Q_{\Phi W} = N^{-1}\Phi^{T} W W^{T} \Phi$. Furthermore, the na\"ive bootstrap estimator results in a sample average of $(\sigma^2_\varepsilon + \sigma^2_\gamma N^{-1}\operatorname{tr}(WW^{T}))\,Q_\Phi^{-1}$.
\end{theorem}

Theorem~\ref{thm:naivefail} states that the na\"ive bootstrap estimator will not produce correct confidence intervals in general. A proof is given in the Appendix. To construct valid confidence intervals, we need to correctly specify $\Phi(W_i, E)$ and estimate both $\sigma^2_\varepsilon$ and $\sigma^2_\gamma$ properly. We suggest the parametric bootstrap procedure summarized in Algorithm~\ref{alg:parboot}.

\begin{algorithm}[t]
\caption{Parametric bootstrap.}\label{alg:parboot}
\KwIn{$B$, the number of bootstrap samples; $Y$, the outcome variable of interest; $\Phi(W,E)$, a function of graph weights $W$ and exposures $E$.}
Regress $Y$ on $\Phi(W, E)$ to estimate $\hat\beta$, with $\hat u$ being the residuals\;
Regress $\hat u$ on $W$ to obtain $\hat\varepsilon$ as residuals\;
$\hat\sigma^2_\varepsilon \leftarrow \hat\varepsilon^{T}\hat\varepsilon / N$\;
$\hat\sigma^2_\gamma \leftarrow (\hat u^{T} \hat u - N\hat\sigma^2_\varepsilon)/\operatorname{tr}(WW^{T})$\;
\For{$b = 1, \ldots, B$}{
  Sample $\gamma^{b} \sim \mathcal{N}(0, \hat\sigma^2_\gamma)$\;
  Sample $\varepsilon^{b} \sim \mathcal{N}(0, \hat\sigma^2_\varepsilon)$\;
  $Y^{b} \leftarrow \Phi(W, E)\hat\beta + W\gamma^{b} + \varepsilon^{b}$\;
  Regress $Y^{b}$ on $\Phi(W, E)$ to obtain $\hat\beta^{b}$\;
}
Use the distribution of $\hat\beta^{b} - \hat\beta$ as an approximation for the distribution of $\hat\beta - \beta$\;
\end{algorithm}

Assuming that outcomes follow the structural form of Equation~\ref{eq:corrmodel}, the parametric bootstrap method in Algorithm~\ref{alg:parboot} recovers the correct distribution of its parameters, which is formalized in the following theorem.

\begin{theorem}[Parametric bootstrap validity]\label{thm:parboot}
The parametric bootstrap procedure outlined in Algorithm~\ref{alg:parboot} leads to valid confidence intervals under the model assumptions of Theorem~\ref{thm:naivefail}.
\end{theorem}

A proof is included in the Appendix. We validate this method empirically in Section~6.3.

\section{Simulation results}
\subsection{Buyer--seller simulation: static bipartite networks}

In this section, we illustrate how the proposed methods perform in a static bipartite network resembling a buyer–seller marketplace. Specifically, we construct a simulation in which diversion units are sellers and outcome units are buyers. We have $M=50$ sellers (or products). Treatment is assigned via an i.i.d.  Bernoulli $(p)$ mechanism with $p=0.5$. In other words, each seller $j$ is independently assigned either treatment $\left(Z_j=1\right)$ or control $\left(Z_j=0\right)$.

 We have $N=1000$ buyers. Each buyer $i$ has an unobserved intention or demand $S_i \sim \operatorname{Poisson}(\lambda)$, here with $\lambda=10$. This $S_i$ drives the number of product connections (edges) that buyer $i$ forms with sellers in the bipartite graph.

For each buyer $i$, the (latent) Poisson variable $S_i$ is first sampled. We then draw the degree $\left|W_{i \cdot}\right|$ of buyer $i$ via Poisson($S_i$). Next, $\left|W_{i \cdot}\right|$ distinct sellers are randomly chosen (without replacement) from $\{1, \ldots, M\}$, and those edges $\left(W_{i j}\right)$ are set to 1; all other edges from buyer $i$ to sellers are 0. Figure 4 shows one such realization of the resulting bipartite network: red nodes are treated sellers, blue nodes are control sellers, and circles around the perimeter represent 50 buyers.

\begin{figure}[htbp]
    \centering
    \includegraphics[width=0.5\textwidth]{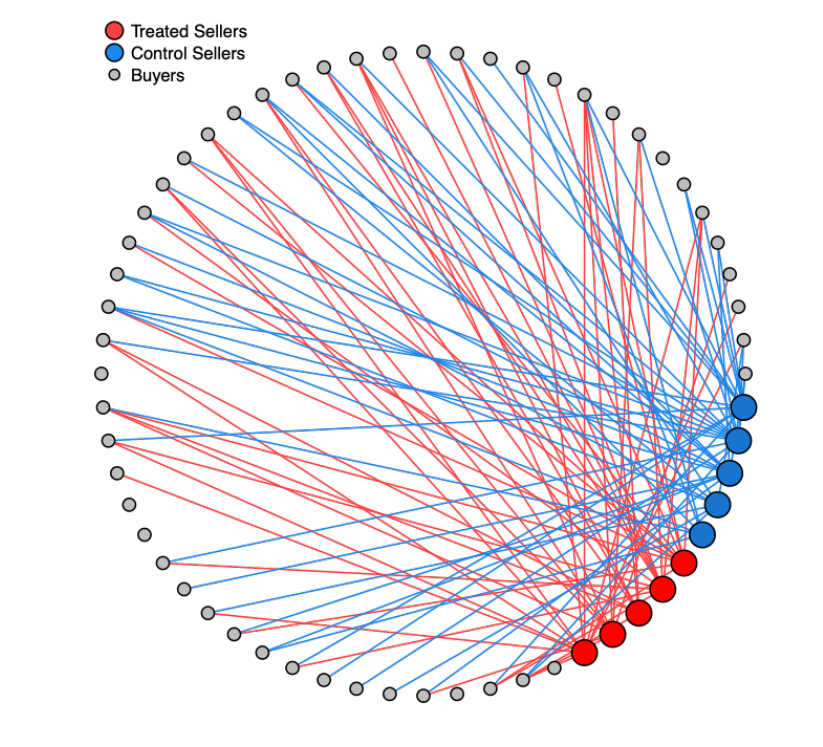}
    \caption{Static Network Simulation: Static Bipartite Network Example}
    \label{fig:network}
\end{figure}

 Once the network is formed, each buyer's exposure  $E_i$ is computed as

$$
E_i=\sum_{j \in\{\text { connected sellers }\}} W_{i j} Z_j .
$$

Because $W_{i j}=1$ whenever buyer $i$ is linked to seller $j, E_i$ simply counts how many of buyer $i$ 's connected sellers are treated.

We specify a "true" exposure effect $\tau=10$ and an "intention" effect $\psi=10$. The potential outcome $Y_i(e)$ (for exposure level $e$ ) depends on both $e$ and the latent demand $S_i$. Concretely,

$$
Y_i(e)=1+\tau e+\psi S_i+\varepsilon_i
$$

where $\varepsilon_i \sim \mathrm{~N}(0,1)$. Although $S_i$ is unobserved in practice, it directly impacts $Y_i$. Figure 5 shows how the observed outcomes $Y_i$ vary with the realized exposures $E_i$.

\begin{figure}[htbp]
    \centering
    \includegraphics[width=0.8\textwidth]{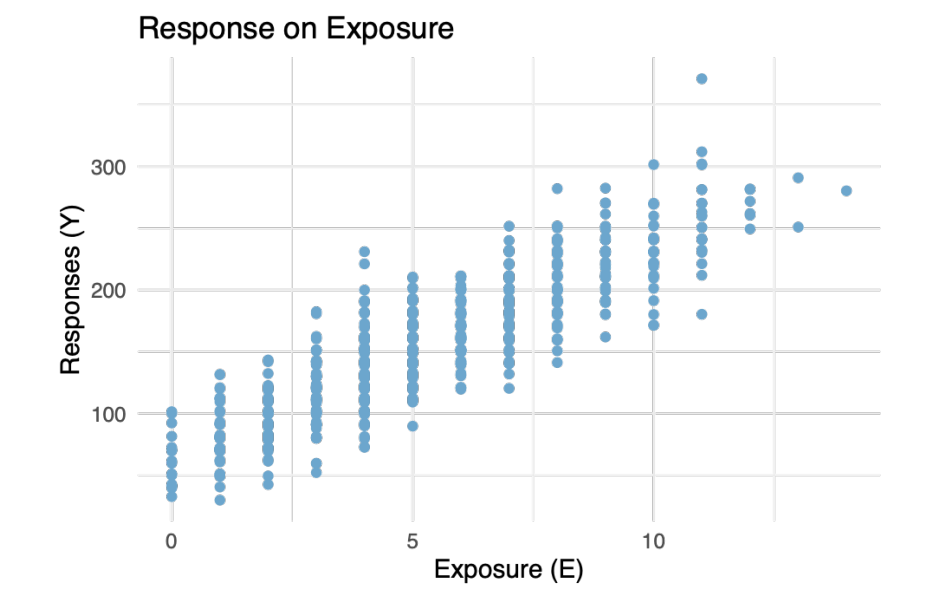}
    \caption{Static Network Simulation: Exposure-Outcome Scatter Plot}
    \label{fig:response}
\end{figure}

 Because $S_i$ is random, the marginal exposure-outcome relationship $\mu(e)$ is obtained by integrating out $S_i$. In this design, we have

$$
\mu(e)=\mathbb{E}\left[Y_i(e)\right]=101+10 e
$$

Hence, the true exposure-response effect is $\tau=10$.
Figure 6 compares point estimates and $90 \%$ confidence intervals for various methods. These include the "baseline" naive regression (which ignores network-related confounding), inverse-propensity weighting (IPW), generalized additive models (GAM) that condition on estimated propensities, and doubly deconfounded (DD) estimators that combine modeling and weighting. Figure 7 shows estimated exposure-response curves alongside the true $\mu(e)$, illustrating how closely each method tracks the correct linear slope of 10.

\begin{figure}[htbp]
    \centering
    \includegraphics[width=0.5\textwidth]{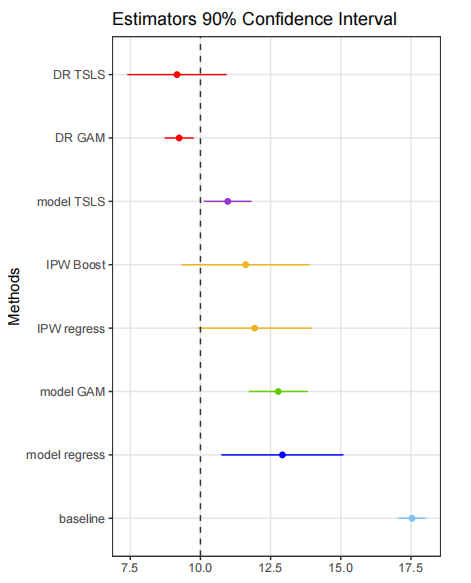}
    \caption{Static Network Simulation: Exposure-Outcome Effect 90\% Confidence Intervals}
    \label{fig:estimators}
\end{figure}

\begin{figure}[htbp]
    \centering
    \includegraphics[width=0.5\textwidth]{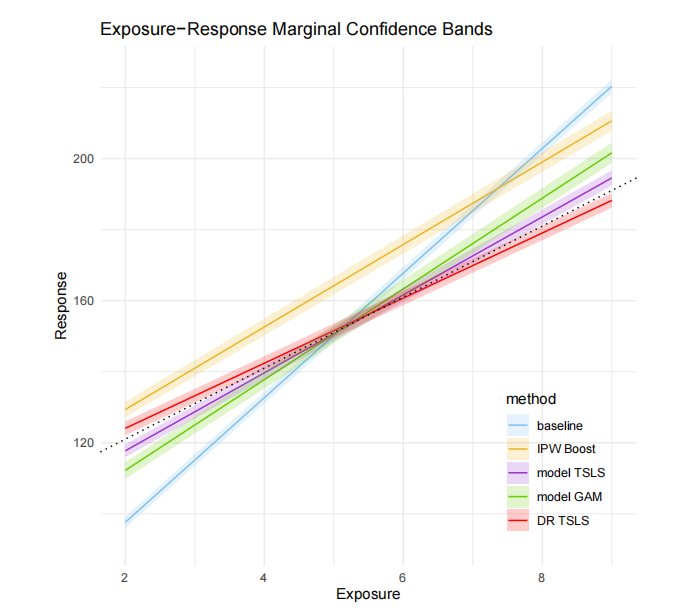}
    \caption{Static Network Simulation: Marginal Exposure-Outcome with 90\% Confidence Bands}
    \label{fig:exposure-response}
\end{figure}

Beyond a single network draw, we also explore how hyper-parameters influence the mean squared error $\operatorname{MSE}(\hat{\tau})$. Specifically, we vary
$M$, the number of sellers $\{50,75,100\}$,
$N$, the number of buyers $\{500,1000,1500\}$,
 $\lambda$, the Poisson mean for buyer demand $\{7,10,13\}$,
 $p$, the Bernoulli treatment probability $\{0.3,0.5,0.7\}$.

For each configuration, 100 independent datasets are generated, yielding 100 estimates $\{\hat{\tau}\}$. We then compute

$$
\operatorname{MSE}(\hat{\tau})=[\operatorname{Bias}(\hat{\tau})]^2+\operatorname{Var}(\hat{\tau})
$$

Here, $\operatorname{Bias}(\hat{\tau})=\mathbb{E}[\hat{\tau}]-\tau$ and $\operatorname{Var}(\hat{\tau})$ is the empirical variance of the estimates.
Table 2 reports the average $\operatorname{MSE}(\hat{\tau})$ across these runs for each method, while Table 3 provides p -values from an ANCOVA that evaluates whether varying each hyper-parameter $(M, N$, $\lambda$, or $p$ ) significantly changes $\operatorname{MSE}(\hat{\tau})$. Several key takeaways emerge.

\begin{table}[htbp]
    \centering
    \includegraphics[width=\textwidth]{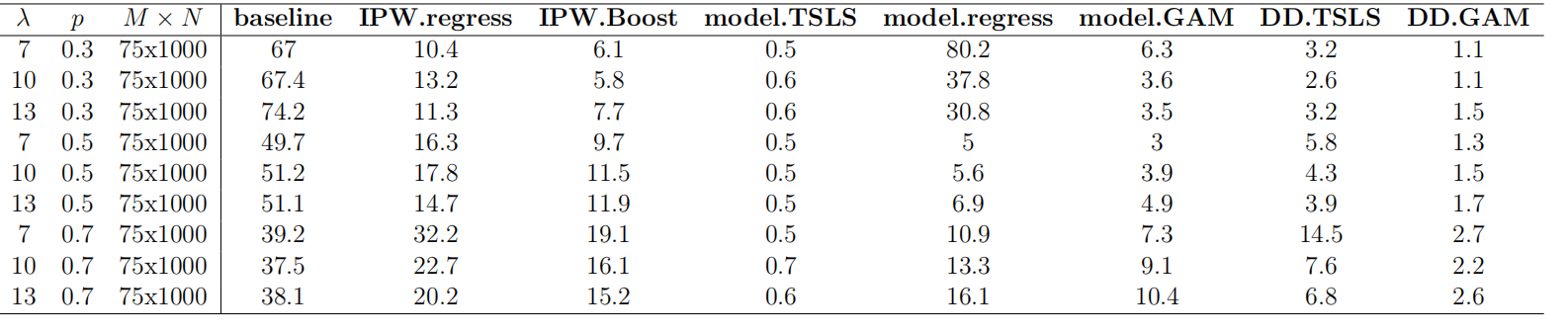}
    \caption{Static Network Batch Simulation: $MSE(\hat{\tau})$ Averages Sample}
    \label{tab:static}
\end{table}

\begin{table}[htbp]
    \centering
    \includegraphics[width=0.5\textwidth]{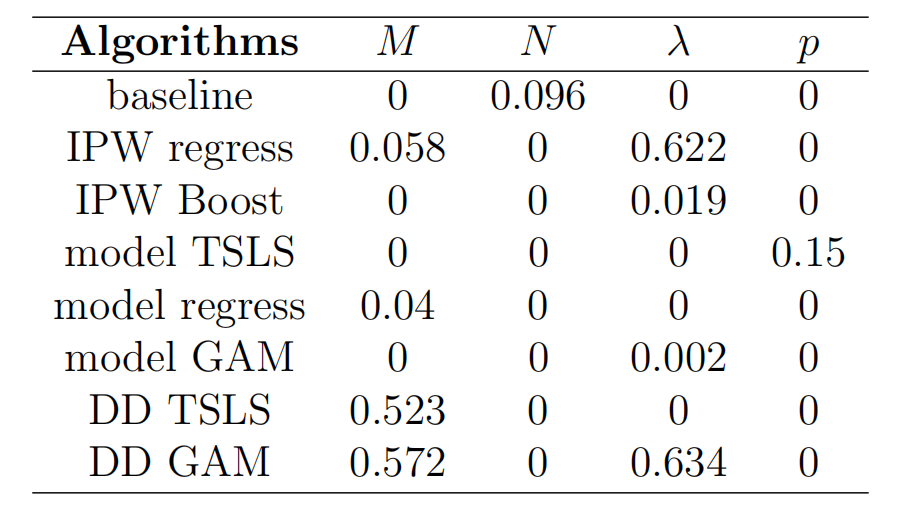}
    \caption{Static Network Batch Simulation: P-values from ANCOVA Analysis Comparing $MSE(\hat{\tau})$ Across Simulation Hyper-parameters for Different Algorithms}
    \label{tab:algo}
\end{table}

First, exposure variability improves estimates. Larger $\lambda$ or higher $p$ both induce more variability in buyer exposures (i.e., more chance of being connected to multiple treated sellers). This greater spread of exposures often reduces MSE because the estimators can better distinguish how outcomes change over a wider range of $E_i$.

Second, the double deconfounded methods demonstrate  stability across different hyper-parameter configurations. Methods that combine propensity weighting with outcome modeling consistently achieve lower MSE, showcasing their double robustness property.

Third, there exists relative sensitivity across algorithms. Some methods are more sensitive to certain hyperparameters than others (e.g., IPW can become unstable if $p$ is small). The $p$-value analysis in Table 3 highlights which parameters ($M, N, \lambda, p$) most strongly affect each algorithm's performance.

Overall, these simulations underscore how bipartite network structure and variability in exposure can materially affect estimation accuracy. By accounting for latent buyer intentions and carefully modeling exposures, the proposed estimators deliver substantial bias reduction relative to naive approaches.

\subsection{Buyer--seller simulation: dynamic bipartite networks}

We now extend the buyer--seller marketplace simulation to a dynamic bipartite network. In this setting, both the demand and the network connections evolve over time, capturing the notion that buyers gradually learn about sellers' products, respond to advertisements, and form new connections as adoption spreads.

We continue to designate $N=1000$ buyers (outcome units) and $M=50$ sellers (diversion units).  As in the static case, each seller $j$ is assigned treatment $Z_j$ via an i.i.d.\ Bernoulli($p=0.5$) design.  The exposure effect $\tau$ and the intention effect $\psi$ are both set to 10.  However, rather than having buyer--seller connections remain fixed, we allow them to evolve across discrete time periods $t=1,\dots,T$, reflecting real-world processes where customer awareness and product demand increase over time.

To model time-varying demand and network formation, we employ a logistic growth function
\[
\eta(t) \;=\; \frac{K}{1+\exp\bigl(-b\,(t-t_0)\bigr)},
\]
where \(K=200\) is the carrying capacity (the largest average degree the network approaches), \(b=0.1\) is the growth rate, and \(t_0=5\) is the time of median growth.  Conceptually, $\eta(t)$ governs how quickly new buyer--seller connections form, mimicking phases of lag, acceleration, rapid adoption, deceleration, and saturation over time.  Figure (panels T1--T5) illustrates snapshots of this bipartite network at five representative stages of its evolution.

At each time $t$, buyer~$i$ forms $\lvert W_{i \cdot}\rvert_t$ new edges to sellers according to a stochastic process with mean~$\eta(t)$.  Let
$\bigl(W_{i\cdot}\bigr)_t$
denote the set of sellers to which buyer~$i$ is connected at time~$t$.  The dynamic exposure of buyer~$i$ at time~$t$ is then
\[
E_{i t} \;=\; \sum_{j \,\in\, (W_{i \cdot})_t} (W_{i j})_t \,Z_j,
\]
which counts how many of buyer~$i$’s current sellers are treated.

We postulate that each buyer’s potential outcome $Y_{i t}(e)$ depends on two drivers: the exposure level $e$ and the buyer’s (latent) time-dependent demand $S_{i t}$.  Thus,  
\[
Y_{i t}\bigl(E_{i t}\bigr)\;\bigm|\;S_{i t}
 \;=\; 1 + \tau\,E_{i t} \;+\;\psi\,S_{i t} \;+\;\varepsilon_{i},
\]
where \(\varepsilon_{i}\sim N(0,1)\).  Over time, $S_{i t}$ also grows with the same or similar S-shaped schedule, reflecting that each buyer’s demand (and hence the inclination to connect) increases as more information accumulates.  

\begin{figure}[t]
    \centering
    \includegraphics[scale=0.5]{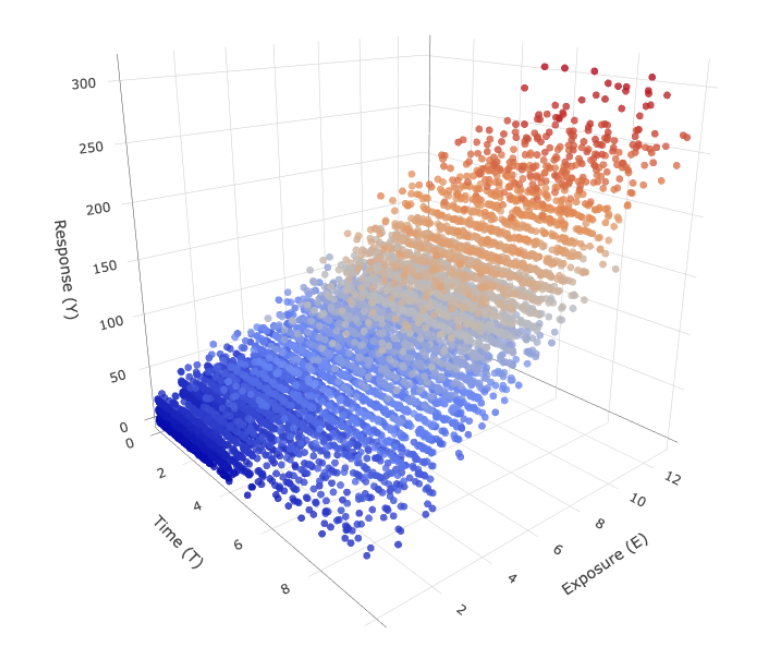}
    \caption{Dynamic Networks Simulation (example run).  A 3D scatter plot of time ($T$), exposure ($E$), and response ($Y$) shows how outcomes rise over time (blue to red gradient) in an S-shaped pattern.}
    \label{fig:3Dscatter-dynamic}
\end{figure}

Figure~\ref{fig:3Dscatter-dynamic} plots one realization of the simulated outcomes over time, exposure, and response.  A distinct S-shaped pattern in the response emerges as $t$ increases, driven by logistic growth in both the demand $S_{i t}$ and the network size.  To capture this visually, we fit a three-parameter logistic curve 
\[
Y(T) \;=\;\frac{\widehat{K}}{1+\exp\bigl[-\widehat{b}\,(T-\widehat{t}_0)\bigr]},
\]
to the (time-aggregated) average responses, yielding estimates 
\(\widehat{K}=150.7,\;\widehat{b}=0.925,\;\widehat{t}_0=4.98\).  As time progresses, not only does the mean outcome level rise, but its variance grows as well, reflecting that some buyers rapidly adopt while others remain slow to learn.

As a baseline, we first consider a na\"ive dynamic regression model:
\[
Y_{i t}\mid E_{i t}
\;\sim\;
N\Bigl(\,\xi_0 \;+\; \xi_1\,E_{i t} \;+\; \sum_{t\in T}\xi_t\,FT_{i t}\;,\;\sigma_y^2\Bigr),
\]
where $FT_{i t}$ is an indicator for time period $t$ (i.e.\ a dynamic intercept).  This approach effectively absorbs time effects into $\sum_{t\in T}\xi_t FT_{i t}$, without assuming any particular parametric form (logistic or otherwise) for how $Y$ grows.  We estimate $\xi_1$ by ordinary least squares with factor-coded time.    Incorporating factor-time intercepts does capture the global S-shaped trajectory better than a purely static model, but it does not deconfound exposure from buyer demand or the time-varying network.  

\begin{figure}[t]
    \centering
    \includegraphics[scale=0.55]{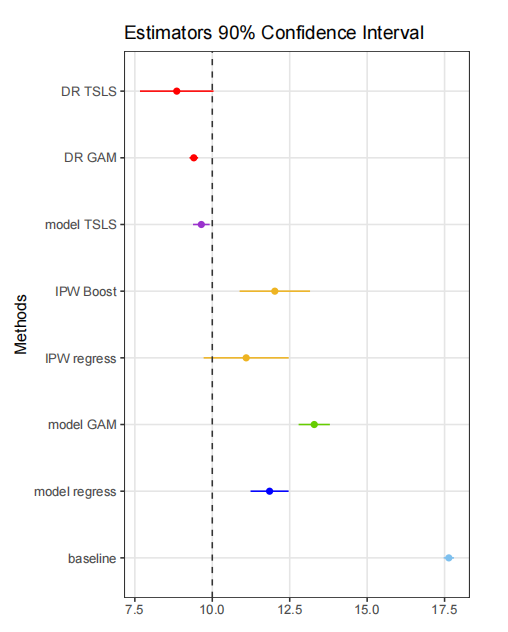}
    \caption{Dynamic Networks Simulation: Comparing point estimates and 90\% confidence intervals for $\hat{\tau}$ under multiple estimation methods, in one representative parameter setting.  The true effect is shown as a vertical dashed line.}
    \label{fig:dyn-exposure-CI}
\end{figure}

We next apply the model-based and weighting-based algorithms  to this dynamic context.  In particular, we revisit TSLS, model regression (GAM), IPW, and double deconfounded methods, now allowing time-varying degrees and logistic growth in the exposure.  
To investigate sensitivity to the growth-rate parameters, we conduct a batch simulation varying:
\[
\begin{aligned}
&p\in\{0.4,\,0.5,\,0.6\},\quad
K\in\{100,\,200,\,300\},\\
&b\in\{0.02,\,0.05,\,0.10\},\quad
t_0\in\{30,\,50,\,70\}.
\end{aligned}
\]
Each configuration is replicated 30 times.  We compute 
\(\mathrm{MSE}(\hat{\tau})=\mathrm{Bias}(\hat{\tau})^2+\mathrm{Var}(\hat{\tau})\)
and compare across algorithms.  Table~\ref{tab:dyn-batch-MSE} displays average MSEs; Figure~\ref{fig:dyn-exposure-CI} shows point estimates and confidence intervals for one representative parameter setting.  Finally, Table~\ref{tab:dyn-ancova} reports $p$-values from an ANCOVA testing how each hyper-parameter affects performance.

\begin{table}[t]
    \centering
    \includegraphics[width=0.9\textwidth]{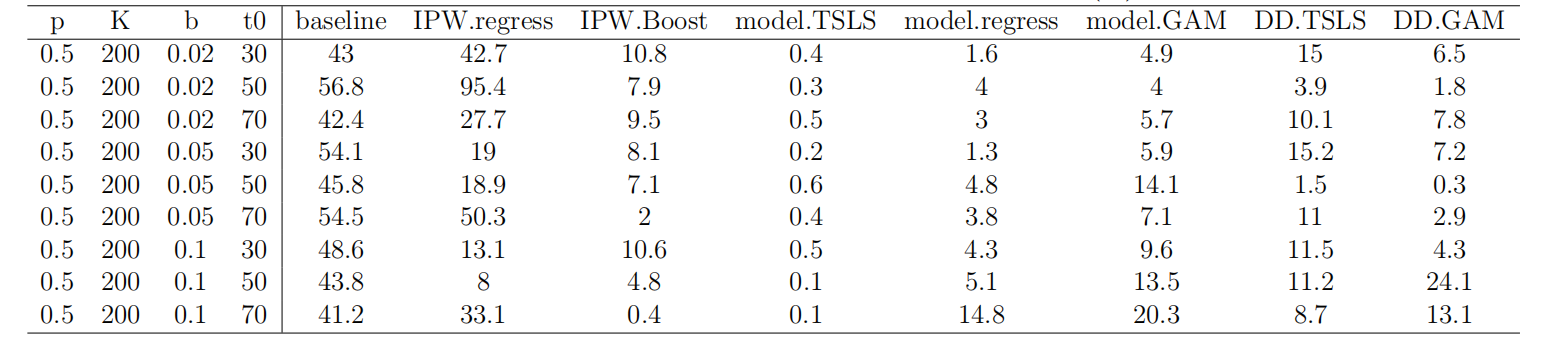}
    \caption{Dynamic Networks Batch Simulation: MSE$(\hat{\tau})$ for various algorithms, averaged over 30 replications, under different combinations of $(p,K,b,t_0)$.}
    \label{tab:dyn-batch-MSE}
\end{table}

\begin{table}[t]
    \centering
    \includegraphics[width=0.5\textwidth]{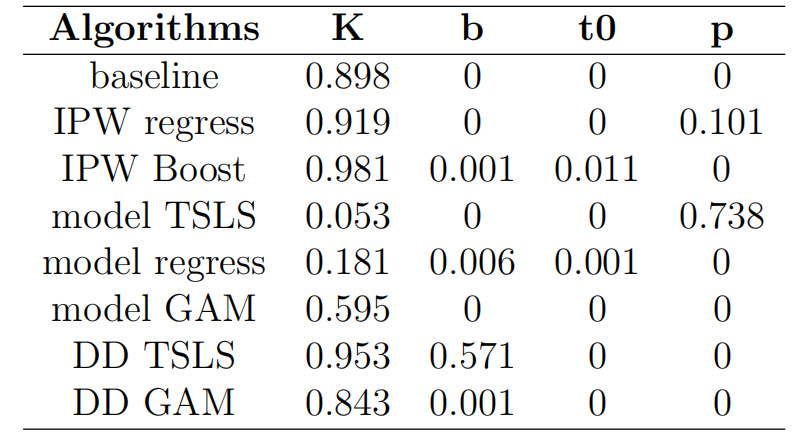}
    \caption{Dynamic Networks Batch Simulation: $p$-values from ANCOVA analysis comparing $\mathrm{MSE}(\hat{\tau})$ across simulation hyper-parameters $(p,K,b,t_0)$.  Low $p$-values highlight sensitivity of each method to a given parameter.}
    \label{tab:dyn-ancova}
\end{table}

In these dynamic simulations, we again observe that increasing exposure variability (e.g.\ via larger $b$ or a more balanced $p$) generally lowers MSE, because the estimator sees a broader range of exposure levels over time.  Most algorithms benefit from time variation so long as the modeling either includes dynamic intercepts (in naive regression) or suitably accounts for time in the exposure equation (e.g.\ TSLS and DD\,TSLS).  

Interestingly, we find that the carrying capacity $K$ (which primarily scales the maximum degree of the bipartite network) matters less than the growth rate $b$ or the shift $t_0$.  Indeed, large~$K$ alone does not necessarily widen the range of exposures across time; it simply raises the eventual ceiling of possible connections.  By contrast, a larger or faster growth rate $b$ produces greater heterogeneity in exposure earlier, which more effectively identifies the slope \(\tau\).  

In summary, introducing temporal dynamics via logistic growth leads to realistic $S$-shaped adoption and learning in a buyer--seller network.  Simulations results underscore the importance of (i) modeling time-varying factors explicitly when the network evolves and (ii) ensuring adequate variability in treatment and connections for reliable causal inference.

\subsection{Coverage under correlated errors}

To validate the variance-estimation results of Section~5, we consider a simulated bipartite graph consisting of $N = 1000$ outcome units and $M = 100$ diversion units. Each outcome unit $i$ is connected to $m_i$ diversion units, where $m_i$ is distributed uniformly over the set of integers from 1 to 10. All weights $W_{ij}$ are set to be equal to $1/m_i$. For the diversion units, the treatment assignments $Z_j$ are chosen to be i.i.d.\ Bernoulli random variables with parameter $p = 1/2$, and we set $\sigma^2_\varepsilon = 0.5$. We construct nominally 95\% confidence intervals using 200 bootstrap samples.

We ran a set of simulations allowing correlated errors and setting $\sigma^2_\gamma = 0.5$. We consider the case of homogeneous treatment effects and compare the na\"ive bootstrap against the parametric bootstrap approach proposed in Section~5. We assume that the functional form of $\Phi(W_i, E_i)$ is known, as discussed in that section, and show that the parametric approach achieves the coverage of 97\%, while the na\"ive approach achieves only 75\% coverage, validating the claim made in Theorem~\ref{thm:parboot}.

\section{Application to the Amazon Pet Supplies data}

We now illustrate our methods on the Amazon Pet Supplies dataset, which was previously analyzed by \citet{pouget2019variance} and \citet{ni2019justifying}.  This dataset, which spans from October 1, 1998, to July 1, 2014, provides detailed product information and user reviews for multiple product categories on Amazon. In our analysis, we focus on the "Pet Supplies" subcategory —one of the largest in the dataset—comprising 740985 unique reviewers and 103288 distinct products.

Table 6 further summarizes basic statistics for both ratings and product costs, while Table 7 provides insight into the distribution of the number of reviews per reviewer (i.e., how many different products a given reviewer rated).

\begin{table}
\centering
\includegraphics[width=0.5\textwidth]{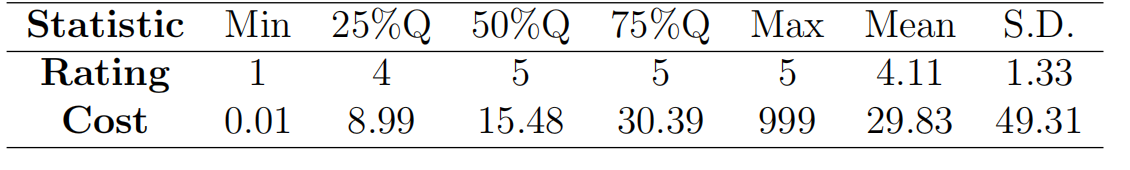}
\caption{Summary Statistics of Amazon Pet Supplies Reviews}
\label{tab:summary-stats}
\end{table}

\begin{table}
\centering
\includegraphics[width=0.5\textwidth]{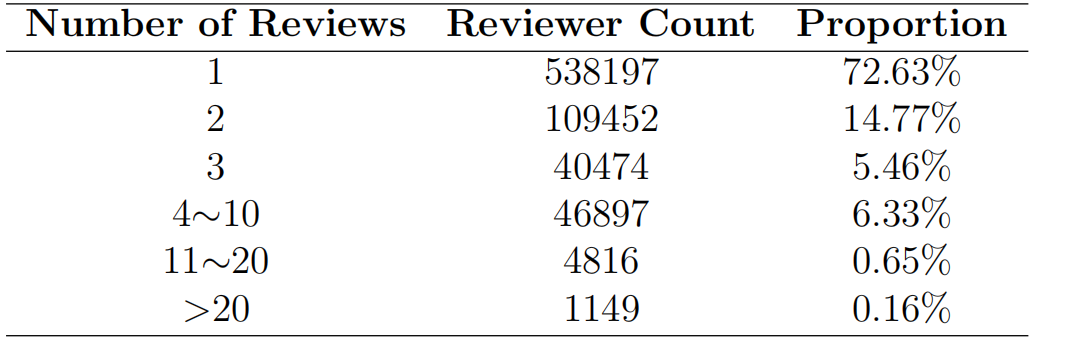}
\caption{Distribution of Reviews per Reviewer}
\label{tab:review-distribution}
\end{table}

We first examine how product rankings may shape customers' purchasing decisions and reviews.
Following earlier bipartite designs, we treat reviewers (buyers) as outcome units and products (items) as diversion units, with the treatment assignment at the product level manifested by each item's ranking. 

Concretely,
We define the outcome of interest for each reviewer as the total amount spent on Pet Supplies products, aggregated across that reviewer's purchase history.
We measure exposure to the ranking system by summing the inverse rank of each product the reviewer purchased. Products with higher (worse) numeric ranks contribute smaller values of 1 / rank; conversely, highly ranked items contribute larger values. Summing these values across all purchased items captures a reviewer's overall "exposure" to higher-ranked products.

Because reviewers often explore multiple Pet Supplies products, we construct a network confounder by summing the number of other items  that each reviewer browsed in this same subcategory. This variable may induce confounding if, for example, reviewers who browse many similar items end up with systematically different purchase behaviors.

The magnitude of these variables varies widely across reviewers: some users buy only one or two items over many years, whereas highly engaged users purchase in bulk and explore dozens of alternatives. To address this heterogeneity, we employ quantile normalization on both the exposure and the network confounder, splitting each into ten quantile-based strata $\left\{E_1, \ldots, E_{10}\right\}$ and $\left\{D_1, \ldots, D_{10}\right\}$, respectively.

Table 6.3 presents estimates of the marginal exposure-cost effect using several of our proposed algorithms, along with baseline methods. Figure 6.2 shows the corresponding 90\% confidence intervals.
The naive "baseline" regression yields an estimate of 4.4193 (S.E. 0.0213 ), which suggests that each additional unit of inverse ranking exposure is associated with a roughly $\$ 4.42$ increase in reviewers' total spending. However, once we adjust for network confounding via IPW, TSLS, GAM, or the doubly deconfounded (DD) methods, the point estimates range between about 7 and 10.5. Most of these lie markedly above the naive estimate, reflecting that ignoring confounding underestimates the true effect of exposure.

\begin{table}[htbp]
    \centering
    \includegraphics[width=0.5\textwidth]{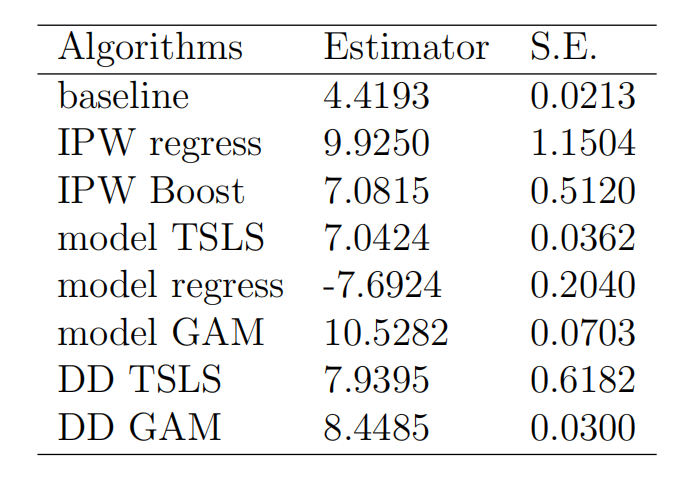}
    \caption{Marginal Exposure-Cost Effect Estimates and Standard Errors for Different Algorithms}
    \label{tab:exposure-cost}
\end{table}

\begin{figure}[htbp]
    \centering
    \includegraphics[width=0.5\textwidth]{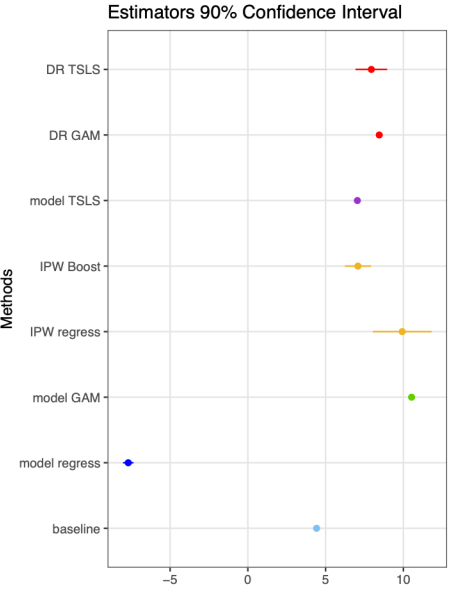}
    \caption{90\% Confidence Intervals for Marginal Exposure-Cost Effect Across Different Algorithms}
    \label{fig:confidence-intervals}
\end{figure}

Interestingly, one method ("model regress") yields a negative estimate of -7.6924 (S.E. 0.2040). We suspect this outlier stems from misspecification in the purely parametric outcome model $\beta(e, r)$. With such a high-dimensional setting, purely parametric assumptions can fail to capture the underlying relationships among rank, cost, and browsing behavior, leading to biased estimates.

Overall, most of the adjusted methods produce positive-and often substantially larger exposure-cost estimates than the baseline. These findings suggest that higher-ranked items in the Pet Supplies category indeed encourage greater spending, but failing to control for each reviewer's browsing intensity can obscure this relationship.

We next turn to the marginal exposure-rating effect, now using the same bipartite structure but defining the outcome of interest as the sum of a reviewer's product ratings. Table 9 summarizes these results. The "baseline" regression  produces a relatively large effect, about 0.8647 (S.E. 0.0015). However, most of the adjusted estimators—particularly IPW, TSLS, GAM, and their double deconfounded variants—now lie between 0.1914 and 0.3954 , all distinctly smaller than the baseline.

\begin{table}[htbp]
    \centering
    \includegraphics[width=0.5\textwidth]{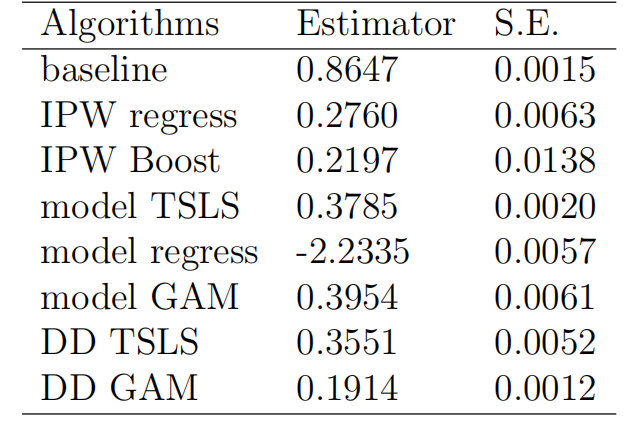}
    \caption{Marginal Exposure-Rating Effect Estimates and Standard Errors for Different Algorithms}
    \label{tab:exposure-rating}
\end{table}

One notable exception arises from the "model regress" approach, which produces an estimate of -2.2335 (S.E.  0.0057). Consistent with our observations above, this starkly negative value likely indicates that the parametric model fails to accommodate the network structure or captures the underlying review/rating process inaccurately.

From a practical standpoint, these results hint that while rank exposure can significantly affect how much a customer spends, it appears less critical in determining how favorably they rate a product. In other words, customers might be nudged by prominent rankings to spend more or to choose certain products, but when writing their reviews, they tend to concentrate on intrinsic product quality—less swayed by a product’s list placement.

Our empirical analysis of the Amazon Pet Supplies data demonstrates the importance of bipartitebased causal methods for mitigating network-induced confounding. For instance, customers who are more inclined to browse additional items (a key confounder) also tend to be more exposed to highly ranked products-complicating any naive attempt to measure "how ranking affects behavior." By using IPW, TSLS, GAM, or their doubly deconfounded counterparts, we see substantial upward corrections in the Exposure-Cost estimates and a generally muted Exposure-Rating effect, underscoring how exposure and reviewers' own behaviors interplay in complex ways.

Additionally, our findings reaffirm the risks of naive parametric modeling in high-dimensional contexts. The stark negative estimates from "model regress" serve as a cautionary tale that conventional regression assumptions may be too restrictive for capturing patterns in large-scale marketplaces. Our more flexible weighting or partially nonparametric methods appear more robust in this regard.

Overall, these results align with the broader conclusion that product rankings can indeed encourage higher spending, but have limited impact on how favorably products are ultimately rated. From a platform-design perspective, such insights help clarify where ranking-based interventions might be most influential in shaping customer behavior (i.e, in purchase decisions), while reminding us that reviewers often focus primarily on product quality when giving feedback.

\section{Concluding remarks}

In this paper, we have emphasized the importance of properly accounting for network-induced biases when designing and analyzing bipartite experiments. Such biases often arise from overlooked exposure distributions that link outcomes on one side of the bipartite graph to treatment assignments on the other. Drawing on foundational work by \citet{imbens2000role}, \citet{hirano2004propensity}, and \citet{robins2000marginal}, we showed how adapting propensity score methodologies to bipartite settings—through modeling, weighting, or doubly deconfounded approaches—can mitigate these biases.

We presented new theoretical insights establishing the unbiasedness of propensity-score-based estimators under standard assumptions, and offered proofs of bounds of the variance of the estimators. Using both synthetic simulations and an observational dataset, we illustrated how conventional, purely model-based methods can yield substantially biased inferences in bipartite designs, whereas the propensity-score-based approaches introduced here markedly reduce that bias.

Looking ahead, there are several valuable directions for future work. First, our framework largely presumes that the underlying network remains fixed and known—subsequent research could investigate estimation and inference in environments where bipartite networks themselves evolve in response to treatment, thereby violating “fixed network” assumptions. Second, many real-world bipartite settings involve correlated errors or non-standard link functions; incorporating flexible models (e.g., Gaussian processes or deep neural networks) may help capture more complex exposure patterns. Finally, our discussion of implementation suggests that high-dimensional, real-time industrial contexts can benefit from these bipartite causal methods. Designing procedures that simultaneously optimize treatment assignment, reduce network confounding, and preserve statistical efficiency remains an exciting and fertile area for future advances in bipartite causal inference.

\nocite{delprete2020continuous,eckles2017design,fatemi2020minimizing,friedman2001elements,galagate2016causal,he2016ups,holtz2025reducing,hong2005effects,hudgens2008toward,johari2022experimental,kempton1997interference,mcauley2015image,ogburn2017causal,rubin1980randomization,saintjacques2019method,saveski2017detecting,struchiner1990behaviour,tchetgen2012causal,toulis2013estimation,viviano2020experimental}
\bibliographystyle{plainnat}
\bibliography{vema_b_references}

\appendix

\section{Additional results}
\subsection{A worked example: na\"ive estimators are biased}

In this section, we show that two different simple estimators, which do not control for the heterogeneity of different outcome units' exposure distributions, are generally biased. For the first estimator, consider using the average of observed outcomes at a given exposure level to estimate the exposure-response function at that exposure level: $\hat\mu(e) = |J(e)|^{-1} \sum_{i \in J(e)} Y_i$, where $J(e) = \{i \in [1, N] : E_i = e\}$ is the set of outcome units with observed exposure $E_i$ equal to $e$. As a slightly more sophisticated estimator, consider running a linear regression of $Y_i$ on $E_i$ (and a constant) and using the regression coefficient as an estimate of the treatment effect $\mu(1) - \mu(0)$. In the following example, we show that both approaches generally produce biased estimates because they do not account for the heterogeneity of exposure distributions and treatment effects.

Suppose we are given a simple bipartite graph with two types of outcome units: outcome units of type S (single) are connected to a single diversion unit and outcome units of type D (double) are connected to exactly two diversion units, such that each outcome unit is connected to its own set of diversion units, each diversion unit being connected to a single outcome unit. To simplify the exposition further, we will assume that the graph weights $\{W_{i\cdot}\}$ of outcome units of type S (resp.\ D) are equal to 1 (resp.\ 1/2), such that the weights corresponding to a given outcome unit always sum to one, and that the two types are present in equal proportions in the graph. Finally, suppose that only units of type D react to treatment. Namely, $\forall e$, $Y_i(e) = 0$ for units of type S and $Y_i(e) = e$ for units of type D.

Assuming a treatment assignment sampled uniformly at random with probability $p = 1/2$, units of type S can receive two levels of exposure with equal probabilities ($E_S = 0$ or 1 with probability 1/2), while units of type D can receive three ($E_D = 0$ or 1 with equal probabilities 1/4, or $E_D = 1/2$ with probability 1/2). The first estimator estimates $\mu(0)$ correctly since $\hat\mu(0) = \mu(0) = 0$, but estimates $\mu(1)$ incorrectly since $\hat\mu(1) = 1/3 < 1/2 = \mu(1)$. The discrepancy occurs because units at exposure level 1 are twice more likely to be of type S than D and not react to treatment, despite both types being equally present in the population. The regression estimator is also biased since $\operatorname{cov}(E_i, Y_i)/\operatorname{var}(E_i) = 1/3 < 1/2 = \mu(1) - \mu(0)$. The fact that these two methods produce identical estimates is purely a coincidence. Their estimates will generally be different since the regression approach accounts for outcome values at all observed levels of exposure while the nonparametric approach depends on $Y_i(0)$ and $Y_i(1)$ only.

To illustrate the merit of the generalized propensity score estimator on this example, we begin by computing the average of potential outcomes at all levels of exposure and propensity score, $\beta(e, r)$:
$$
\beta(e, r) =
\begin{cases}
0, & (e, r) \in \{(0, 1/2) \cup (1, 1/2) \cup (0, 1/4)\} \\
1/2, & (e, r) = (1/2, 1/2) \\
1, & (e, r) = (1, 1/4).
\end{cases}
$$
To estimate the exposure response curve at 0 and 1, we compute the average of $\beta$ at $e = 0$ and $e = 1$, making sure to use the propensity score of each outcome unit for the imputed exposure level, as opposed to the propensity score for their observed exposure level. Units of type S (resp.\ type D) have the propensity score of 1/2 (resp.\ 1/4) at the exposure levels $e \in \{0, 1\}$, leading to $\hat\mu(e) = 1/2 \cdot \beta(e, 1/2) + 1/2 \cdot \beta(e, 1/4)$ since each type is present in equal proportions. It follows that $\hat\mu(e)$ is equal to 0 if $e = 0$ and 1/2 if $e = 1$, in line with the true exposure response function $\mu(e) = e/2$.

To illustrate the statement of Theorem~\ref{thm:weighting} let us consider, for example, $e = 1$. For units of type S, $D_i(1) = 1$ implies that $Y_i = 0$ and $r(1, W_i) = 1/2$. In the overall population units of type S with $D_i(1) = 1$ represent half of all type S units, or a quarter of all units. For units of type D, $D_i(1) = 1$ implies that $Y_i = 1$, $r(1, W_i) = 1/4$, and these units represent one quarter of all type D units, or one-eighth of all units in the population. For any unit with $D_i(1) = 0$ its contribution to the expectation in the right-hand side of the equation in Theorem~\ref{thm:weighting} is zero. As a result, we estimate $\mu(1)$ as $(1/(1/4)) \cdot 1/8 = 1/2$, which is indeed the case.

\subsection{Sufficient conditions for the unbiasedness of na\"ive estimators}

We present two results that provide rather strong sufficient conditions for the unbiasedness of na\"ive estimators.

\begin{proposition}\label{prop:naiveavg}
Suppose that Assumptions \ref{ass:fixedweights} and \ref{ass:weak} hold. If for some $e \in [0,1]$, $\mathbb{E}[Y_i(e) \mid W] = \mu(e)$, then
$$
\mathbb{E}\left[\frac{1}{|J(e)|} \sum_{i \in J(e)} Y_i\right] = \mu(e).
$$
\end{proposition}

In other words, if potential outcomes are the same in expectation regardless of the graph weights, then averaging the outcomes observed at a given exposure level produces an unbiased point estimate of the exposure-response curve. If, in addition to the assumptions in Proposition~\ref{prop:naiveavg}, $\mathbb{E}[Y_i(e)] = \alpha + \beta e$ and strong unconfoundedness holds, the na\"ive regression produces an unbiased estimate too.

\begin{proposition}\label{prop:naiveols}
Suppose that Assumptions \ref{ass:fixedweights} and \ref{ass:weak} hold. If for all $e \in [0,1]$, $\mathbb{E}[Y_i(e) \mid W] = \alpha + \beta e$, then $\hat\Delta = \hat\beta_{\mathrm{OLS}}$, where $\hat\beta_{\mathrm{OLS}}$ is the slope estimate from the regression of $Y_i$ on $E_i$, is an unbiased estimate of $\Delta = \mu(1) - \mu(0)$.
\end{proposition}

\subsection{Details on the unbiased estimator with proper coverage}

Assume that we are interested in finding $\mu(e)$ at a finite number of potential exposures $\{e_1, \ldots, e_R\}$. Let
$$
\Phi(W, E) = \left[\frac{D(e_1)}{\sqrt{r(e_1, W)}}, \ldots, \frac{D(e_R)}{\sqrt{r(e_R, W)}}\right],
$$
and $\tilde{Y} = Y/\sqrt{r(E, W)}$. Then the regression coefficient in front of $D(e_1)/\sqrt{r(e_1, W)}$ from the regression of $\tilde{Y}$ on $\Phi(W, E)$ is given by the expression:
$$
\hat\beta_r = \frac{\sum_{i=1}^{N} \dfrac{Y_i D_{ir}}{r(E_i, W_i)}}{\sum_{i=1}^{N} \dfrac{D_{ir}}{r(E_i, W_i)}}.
$$
As $N \to \infty$, by the law of large numbers:
$$
\frac{1}{N} \sum_{i=1}^{N} \frac{D_{ir}}{r(E_i, W_i)} \to_{a.s.} \mathbb{E}\left[\frac{D_r}{r(E, W)} \,\middle|\, \nu\right] = r(e_r, W)\, \mathbb{E}\left[\frac{1}{r(e_r, W)} \,\middle|\, \nu\right],
$$
which is equal to 1. Hence, $\hat\beta_r \to_{a.s.} \frac{1}{N} \sum_{i=1}^{N} \frac{Y_i D_{ir}}{r(E_i, W_i)}$, which was shown to be an unbiased and consistent estimator of $\mu(e_r)$. Theorem~\ref{thm:naiveboot} can then be used for constructing confidence intervals around $\hat\beta$.

\subsection{Formal construction of the data generating process for variance estimation}

In this section, we detail a formal construction of the data generating process that is used to prove the results of Section~5, namely Theorems \ref{thm:naiveboot} and \ref{thm:naivefail}. Each outcome unit's outcome is considered as a realized sequence of $(\Gamma_k, u_k, Z_k, W_k)_{k=1}^{\infty}$, where the set of possible sequences is denoted as $O$ and
\begin{itemize}
\item $u_k$ is a vector of length $N_k$ of unobserved shocks (errors);
\item $\Gamma_k$ is a sequence of graphs with an arbitrary number of diversion units and $N_k$ outcome units;
\item $W_k$ are the corresponding weights of the graph;
\item $Z_k$ is a realization of treatments over the diversion units of graph $\Gamma_k$.
\end{itemize}
There is a sequence of underlying $\sigma$-algebras $S_0 \subseteq S_1 \subseteq S_2 \subseteq \ldots$ over $O$ with the property that $(\Gamma_k, u_k, Z_k, W_k)$ is measurable with respect to $S_k$. $P_0, P_1, \ldots$ are the measures over those $\sigma$-algebras. Note that $\Gamma_k$, $W_k$, and $Z_k$ uniquely define the exposure generating process $E_k$. A few examples below show how this formal construction can be applied rigorously.
\begin{itemize}
\item \textit{Standard treatment effects.} In this case $\Gamma_k$ is a bipartite graph of $k$ diversion units and $k$ outcome units with unweighted edges. $u_k = \{u_1, \ldots, u_k\}$ are i.i.d.\ random variables drawn from a distribution $F_u$ with mean 0 and finite $\sigma^2_u$. $S_k$ is the product $\sigma$-algebra, and $P_k$ is the product measure. $S_0 = \{\emptyset, O\}$.
\item \textit{Weakly dependent blocks.} $\Gamma_k$ is now a graph of $k$ blocks, with edges between each pair of blocks. $W_k$ denote the weights of the inner links and between-links, with the property that the ratio of between-weights to within-weights converges to 0 a.s. In this example, $u_k = \{u_1, \ldots, u_k\}$ are i.i.d.\ random variables drawn from a distribution $F_u$ with mean 0 and finite $\sigma^2_u$. $S_k$ is the product $\sigma$-algebra, and $P_k$ is the joint distribution satisfying the property that the marginal distribution of observing a particular set of blocks (i.e., integrating out $W_k$) results in the product measure over blocks. $S_0 = \{\emptyset, O\}$.
\item \textit{Influential units.} For this specific setting, we consider non-trivial $S_0$ $\sigma$-algebras. For example, the graph contains a single influential unit, with all other units being like the standard treatment effects example. In that case, we have $S_0 = \{\emptyset, 0, 1, \{0,1\}\}$, indicating events of the influential unit being treated or not. The rest of the construction is the same as in the standard treatment example with the exception that $\sigma$-algebras and measures now have to be cross-producted with $S_0$ and the measure over $S_0$, respectively. Note that in this construction we have for any $k$:
$$
P(E_1 \geq 0.5, \ldots, E_k \geq 0.5) = P(\text{inf.\ unit is treated}) = P_k(\{Z_0 = 1\}) = p \nrightarrow 0,
$$
even though all units have independent realizations of $u$'s and $Z$'s.
\item \textit{AR-1 process.} Suppose that $\Gamma$ enumerates both types of units and the link between an outcome and diversion unit exists iff the number of the outcome unit is equal to the number of the diversion unit or greater than that number by exactly one. In this scenario, the path-based distance between any two outcome units $i$ and $j$ is equal to $2|i - j|$. Setting $\operatorname{cov}(u_i, u_j) = \rho^{\mathrm{dist}(i,j)}$ results in the AR-1 process.
\item \textit{Clusters.} Diversion units are a union of two subsets $C$ and $I$. Each of the outcome units is connected to one unit in $C$ and one unit in $I$. $\operatorname{cov}(u_i, u_j) = v_0 \mathbb{I}[c_i = c_j]$.
\end{itemize}

\section{Proofs}

\begin{proof}[Proof of Lemma~1]
First,

$$
\begin{aligned}
\operatorname{Pr}(\mathbb{I}(e)=1 \mid W, r(e, W)) & =\mathbb{E}[\mathbb{I}(e) \mid W, r(e, W)] \\
& =\mathbb{E}[\mathbb{I}(e) \mid W] \\
& =r(e, W)
\end{aligned}
$$

because by definition, $r(e, W)=\mathbb{E}[\mathbb{I}(e) \mid W]$.
Second,

$$
\begin{aligned}
\operatorname{Pr}(\mathbb{I}(e)=1 \mid r(e, W)) & =\mathbb{E}[\mathbb{I}(e) \mid r(e, W)] \\
& =\mathbb{E}[\mathbb{E}[\mathbb{I}(e) \mid W, r(e, W)] \mid r(e, W)] \\
& =\mathbb{E}[r(e, W) \mid r(e, W)] \\
& =r(e, W)
\end{aligned}
$$

Hence, $\operatorname{Pr}(\mathbb{I}(e)=1 \mid W, r(e, W))=\operatorname{Pr}(\mathbb{I}(e)=1 \mid r(e, W))$ and conditionally on $r(e, W)$, the treatment indicator $\mathbb{I}(e)$ and the pre-treatment variables $W$ are independent. 
\end{proof}

\begin{proof}[Proof of Lemma~2]
First,

$$
\begin{aligned}
\operatorname{Pr}(\mathbb{I}(e)=1 \mid Y(e), r(e, W)) & =\mathbb{E}[\mathbb{I}(e) \mid Y(e), r(e, W)] \\
& =\mathbb{E}[\mathbb{E}[\mathbb{I}(e) \mid Y(e), W, r(e, W)] \mid Y(e), r(e, W)] \\
& =\mathbb{E}[r(e, W) \mid Y(e), r(e, W)] \\
& =r(e, W)
\end{aligned}
$$

Second, as shown before in the proof for Lemma 1,

$$
\operatorname{Pr}(\mathbb{I}(e)=1 \mid r(e, W))=r(e, W)
$$

Hence,

$$
\operatorname{Pr}(\mathbb{I}(e)=1 \mid Y(e), r(e, W))=\operatorname{Pr}(\mathbb{I}(e)=1 \mid r(e, W))
$$

and conditionally on $r(e, W)$ the treatment indicator $\mathbb{I}(e)$ and the potential outcome $Y(e)$ are independent. 

\end{proof}

\begin{proof}[Proof of Theorem~1]
Under Assumption 1 and Assumption 2, then: Part (i). First,

$$
\begin{aligned}
\mathbb{E}[Y \mid E=e, r(E, W)=r] & =\mathbb{E}[Y(e) \mid E=e, r(E, W)=r] \\
& =\mathbb{E}[Y(e) \mid E=e ; r(e, W)=r] \\
& =\mathbb{E}[Y(e) \mid \mathbb{I}(e)=1 ; r(e, W)=r]
\end{aligned}
$$

which by unconfoundedness is equal to

$$
\mathbb{E}[Y(e) \mid r(e, W)=r]
$$

Part (ii) follows directly by applying iterated expectations.

\end{proof}

\begin{proof}[Proof of Theorem~2]
First, rewrite the expectation as an iterated expectation with the inner expectation conditional on $W$ :

$$
\mathbb{E}\left[\frac{Y \cdot \mathbb{I}(e)}{r(E, W)}\right]=\mathbb{E}\left[\mathbb{E}\left[\left.\frac{Y \cdot \mathbb{I}(e)}{r(E, W)} \right\rvert\, W\right]\right]
$$

Next, by conditioning on $\mathbb{I}(e)=1$ and multiplying by the probability of $\mathbb{I}(e)=1$ this is equal to

$$
\mathbb{E}\left[\mathbb{E}\left[\left.\frac{Y}{r(E, W)} \right\rvert\, \mathbb{I}(e)=1, W\right] \cdot \operatorname{Pr}(\mathbb{I}(e)=1 \mid W)\right]
$$

Conditional on $\mathbb{I}(e)=1, Y=Y(e)$ and $r(E, W)=r(e, W)$, so this can be rewritten as:

$$
\mathbb{E}\left[\mathbb{E}\left[\left.\frac{Y(e)}{r(e, W)} \right\rvert\, \mathbb{I}(e)=1, W\right] \cdot \operatorname{Pr}(\mathbb{I}(e)=1 \mid W)\right]
$$

Now the conditioning on $\mathbb{I}(e)$ is irrelevant by the weak unconfoundedness assumption, so this is equal to:

$$
\mathbb{E}\left[\mathbb{E}\left[\left.\frac{Y(e)}{r(e, W)} \right\rvert\, W\right] \cdot r(e, W)\right]=\mathbb{E}[\mathbb{E}[Y(e) \mid W]]=\mathbb{E}[Y(e)]
$$

\end{proof}

\begin{proof}[Proof of Lemma~3]
$$
\begin{aligned}
& \operatorname{Pr}(\mathbb{I}(e)=1 \mid \beta(e, r), r(e, W))=\mathbb{E}[\mathbb{I}(e) \mid \beta(e, r), r(e, W)] \\
&=\mathbb{E}[\mathbb{E}[\mathbb{I}(e) \mid \beta(e, r), W, r(e, W)] \mid \beta(e, r), r(e, W)] \\
&=\mathbb{E}[r(e, W) \mid \beta(e, r), r(e, W)] \\
&=r(e, W) \\
& \operatorname{Pr}(\mathbb{I}(e)=1 \mid r(e, W))=r(e, W) .
\end{aligned}
$$

Hence,

$$
\operatorname{Pr}(\mathbb{I}(e)=1 \mid \beta(e, r), r(e, W))=\operatorname{Pr}(\mathbb{I}(e)=1 \mid r(e, W)),
$$

and conditionally on $r(e, W)$ the treatment indicator $\mathbb{I}(e)$ and the potential outcome model $\beta(e, r)$ are independent.
\end{proof}

\begin{proof}[Proof of Theorem~3]
$$
\mathbb{E}\left[\frac{\beta(e, r) \cdot \mathbb{I}(e)}{r(E, W)}\right]=\mathbb{E}\left[\mathbb{E}\left[\left.\frac{\beta(e, r) \cdot \mathbb{I}(e)}{r(E, W)} \right\rvert\, W\right]\right]
$$

Next, by conditioning on $\mathbb{I}(e)=1$ and multiplying by the probability of $\mathbb{I}(e)=1$.

$$
\mathbb{E}\left[\mathbb{E}\left[\left.\frac{\beta(e, r)}{r(E, W)} \right\rvert\, \mathbb{I}(e)=1, W\right] \cdot \operatorname{Pr}(\mathbb{I}(e)=1 \mid W)\right]
$$

Based on the Lemma 3, Conditional on given strata $\mathbb{I}(e)=1, \beta(e, r)$ and $r(E, W)=$ $r(e, W)$, then rewritten as:

$$
\mathbb{E}\left[\mathbb{E}\left[\left.\frac{\beta(e, r)}{r(e, W)} \right\rvert\, W\right] \cdot \operatorname{Pr}(\mathbb{I}(e)=1 \mid W)\right]
$$

Now the conditioning on $\mathbb{I}(e)$ is irrelevant by the weak unconfoundedness assumption, so this is equal to:

$$
\mathbb{E}\left[\mathbb{E}\left[\left.\frac{\beta(e, r)}{r(e, W)} \right\rvert\, W\right] \cdot r(e, W)\right]=\mathbb{E}[\mathbb{E}[\beta(e, r) \mid W]]
$$

We know $\beta(e, r)$ is the outcome model function for the outcome $Y(e)$. By Lemma 1 , given the network information $W$, then $\mathbb{E}[\beta(e, r) \mid W]=Y(e)$. Then above double expectation is equivalent to:

$$
\mathbb{E}[\mathbb{E}[\beta(e, r) \mid W]]=\mathbb{E}[Y(e)]=\mu(e)
$$

\end{proof}

\begin{proof}[Proof of the upper bounds on the variance of $\hat{\mu}(e)$]
Define the summand:
$$
X_i=\frac{1\left(E_i=e\right) Y_i}{r\left(e, W_i\right)}
$$
Then
$$
\hat{\mu}(e)=\frac{1}{n} \sum_{i=1}^n X_i .
$$

Under an assumption of (at least) no correlation among $\left\{X_i\right\}_{i=1}^n$ (for instance, i.i.d. data):
$$
\operatorname{Var}(\hat{\mu}(e))=\operatorname{Var}\left(\frac{1}{n} \sum_{i=1}^n X_i\right)=\frac{1}{n^2} \operatorname{Var}\left(\sum_{i=1}^n X_i\right)=\frac{1}{n^2} \sum_{i=1}^n \operatorname{Var}\left(X_i\right)
$$

Hence,
$$
\operatorname{Var}(\hat{\mu}(e)) \leq \frac{1}{n^2} \sum_{i=1}^n \mathbb{E}\left[X_i^2\right]
$$
because $\operatorname{Var}\left(X_i\right) \leq \mathbb{E}\left[X_i^2\right]$.

We impose two common bounding assumptions:
1. Bounded outcome: $\left|Y_i\right| \leq M$ almost surely, for some finite $M>0$.
2. Positivity: $r\left(e, W_i\right) \geq \delta$ almost surely, for some $\delta>0$, which prevents division by zero or tiny probabilities.

Under these assumptions:
$$
X_i^2=\left(\frac{1\left(E_i=e\right) Y_i}{r\left(e, W_i\right)}\right)^2=\frac{1\left(E_i=e\right) Y_i^2}{r\left(e, W_i\right)^2} \leq \frac{1\left(E_i=e\right) M^2}{\delta^2}
$$

Thus
$$
\mathbb{E}\left[X_i^2\right] \leq \frac{M^2}{\delta^2} \mathbb{E}\left[1\left(E_i=e\right)\right]=\frac{M^2}{\delta^2} \operatorname{Pr}\left(E_i=e\right)
$$

Summing over $i=1, \ldots, n$ and dividing by $n^2$ :
$$
\frac{1}{n^2} \sum_{i=1}^n \mathbb{E}\left[X_i^2\right] \leq \frac{1}{n^2} \sum_{i=1}^n \frac{M^2}{\delta^2} \operatorname{Pr}\left(E_i=e\right)=\frac{M^2}{\delta^2} \frac{\operatorname{Pr}(E=e)}{n}
$$
where $\operatorname{Pr}(E=e)$ (the marginal probability that a given unit has exposure $e$ ) is the same for all $i$ if they are i.i.d.

Therefore,
$$
\operatorname{Var}(\hat{\mu}(e)) \leq \frac{M^2}{\delta^2} \frac{\operatorname{Pr}(E=e)}{n}
$$

Since $\operatorname{Pr}(E=e) \leq 1$, we obtain the simpler bound:
$$
\operatorname{Var}(\hat{\mu}(e)) \leq \frac{M^2}{n \delta^2}
$$
\end{proof}

\begin{proof}[Proof of Proposition~\ref{prop:naiveavg}]
In this proof, as well as the next one, we slightly abuse the notation by using $W$ to refer to all of the graph weights, not just those corresponding to outcome unit $i$. As $i \in J(e)$ if and only if $D_i(e) = 1$, we can write
$$
\begin{aligned}
\mathbb{E}\left[\frac{1}{|J(e)|} \sum_{i \in J(e)} Y_i\right]
&= \mathbb{E}\left[\frac{1}{\sum_i D_i(e)} \sum_{i=1}^{N} Y_i D_i(e)\right] \\
&= \mathbb{E}\left[\sum_{i=1}^{N} \mathbb{E}\left[\frac{D_i(e)}{\sum_i D_i(e)}\, Y_i(e) \,\middle|\, W\right]\right] \\
&= \mathbb{E}\left[\sum_{i=1}^{N} \mathbb{E}\left[\frac{D_i(e)}{\sum_i D_i(e)} \,\middle|\, W\right] \times \mathbb{E}\left[Y_i(e) \mid W\right]\right] \\
&= \mu(e),
\end{aligned}
$$
where the second to last equality follows from weak unconfoundedness and the last equality follows from $\mathbb{E}[Y_i(e) \mid W] = \mathbb{E}[Y_i(e)] = \mu(e)$.
\end{proof}

\begin{proof}[Proof of Proposition~\ref{prop:naiveols}]
For the na\"ive regression estimator $\hat\beta_{\mathrm{OLS}}$ we have
$$
\mathbb{E}\left[\hat\beta_{\mathrm{OLS}}\right] = \frac{\Cov(Y_i, E_i)}{\Var(E_i)},
$$
where
$$
\begin{aligned}
\Cov(Y_i, E_i) &= \mathbb{E}[Y_i E_i] - \mathbb{E}[Y_i]\mathbb{E}[E_i] \\
&= \mathbb{E}\left[E_i \mathbb{E}[Y_i \mid W]\right] - \alpha \mathbb{E}[E_i] - \beta \left(\mathbb{E}[E_i]\right)^2 \\
&= \mathbb{E}\left[E_i (\alpha + \beta E_i)\right] - \alpha \mathbb{E}[E_i] - \beta \left(\mathbb{E}[E_i]\right)^2 \\
&= \beta \cdot \Var(E_i).
\end{aligned}
$$
The second and third equalities follow from the fact that $\mathbb{E}[Y_i(e) \mid W] = \alpha + \beta e$ for all $e \in [0,1]$ and strong unconfoundedness. Consequently, $\mathbb{E}[\hat\beta_{\mathrm{OLS}}] = \beta = \mu(1) - \mu(0) = \Delta$.
\end{proof}

\begin{assumption}\label{ass:cdf}
The cumulative distribution function of $\Phi(W, E)$ produced by the data generating process converges almost surely to $F^{\Phi}(\cdot\,; \nu)$, where $\nu$ is $S_0$-measurable.
\end{assumption}

\begin{proof}[Proof of Theorem~\ref{thm:naiveboot}]
Conditional on any realization of the data, we can write:
$$
\sqrt{N}(\hat\beta - \beta) = \left(\frac{1}{N}\sum_{i=1}^{N} \Phi(W_i, E_i)\Phi(W_i, E_i)'\right)^{-1} \times \left(\frac{1}{\sqrt{N}}\sum_{i=1}^{N} u_i \Phi(W_i, E_i)\right).
$$
By Assumption~\ref{ass:cdf}, the first term converges a.s.\ to some $Q_\Phi(\nu)^{-1}$ and the second term converges in distribution to $\mathcal{N}(0, \sigma^2_u Q_\Phi(\nu))$.\footnote{Independence of the $u_i$'s implies that only the products of $i$ with $i$ survive in the variance term.} As a consequence, $\sqrt{N}(\hat\beta - \beta) \mid \nu \to_{d} \mathcal{N}(0, \sigma^2_u Q_\Phi(\nu)^{-1})$.

Let us calculate the asymptotic intervals for the na\"ive bootstrap and the na\"ive normal approximation. Standard reasoning implies that conditional on $\nu$ both procedures asymptotically approximate $\sqrt{N}(\hat\beta - \beta)$ as $\mathcal{N}(0, \sigma^2_u Q_\Phi(\nu)^{-1})$, which is exactly the same form as the correct distribution. Hence, in the limit we have $P(\beta \in C_\alpha \mid \nu) = 1 - \alpha$ for almost any $\nu$. It then follows that:
$$
P(\beta \in C_\alpha) = \mathbb{E}[P(\beta \in C_\alpha \mid \nu)] \to_{a.s.} \mathbb{E}[1 - \alpha] = 1 - \alpha.
$$
\end{proof}

\begin{assumption}\label{ass:qphiw}
$\frac{1}{N}\Phi' W W' \Phi \to_{a.s.} Q_{\Phi W}(\nu)$ with $\nu$ being $S_0$-measurable.
\end{assumption}

\begin{proof}[Proof of Theorem~\ref{thm:naivefail}]
In this case, we can denote $u_i = \sum_{j=1}^{K} w_{ij}\gamma_j + \varepsilon_i$ and see that:
$$
\mathbb{E}[u \mid E, W] = 0 \quad \text{and} \quad \Var(u \mid E, W) = \sigma^2_\varepsilon I_N + \sigma^2_\gamma W W'.
$$
As a result, $\hat\beta = (\Phi'\Phi)^{-1}\Phi' Y$ is an unbiased estimator of $\beta$ conditional on any realization of $E, W$. Consequently:
$$
\sqrt{N}(\hat\beta - \beta) = \left(\frac{1}{N}\Phi'\Phi\right)^{-1} \left(\frac{1}{\sqrt{N}}\Phi' u\right).
$$
Recall that under Assumption~\ref{ass:cdf}, $\frac{1}{N}\Phi'\Phi \to_{a.s.} Q_\Phi(\nu)$. The variance of $\sqrt{N}(\hat\beta - \beta)$ is given by the expression:
$$
\begin{aligned}
\Var\left(\sqrt{N}(\hat\beta - \beta) \,\middle|\, E, W\right)
&= \mathbb{E}\left[\left(\tfrac{1}{N}\Phi'\Phi\right)^{-1} \left(\tfrac{1}{N}\Phi' u u' \Phi\right) \left(\tfrac{1}{N}\Phi'\Phi\right)^{-1} \,\middle|\, E, W\right] \\
&= \left(\tfrac{1}{N}\Phi'\Phi\right)^{-1} \left(\tfrac{1}{N}\Phi'\,\mathbb{E}\left[u u' \mid E, W\right]\Phi\right) \left(\tfrac{1}{N}\Phi'\Phi\right)^{-1} \\
&= \left(\tfrac{1}{N}\Phi'\Phi\right)^{-1} \left(\tfrac{1}{N}\Phi'\left(\sigma^2_\varepsilon I_N + \sigma^2_\gamma W W'\right)\Phi\right) \left(\tfrac{1}{N}\Phi'\Phi\right)^{-1} \\
&= \sigma^2_\varepsilon \left(\tfrac{1}{N}\Phi'\Phi\right)^{-1} + \sigma^2_\gamma \left(\tfrac{1}{N}\Phi'\Phi\right)^{-1} \left(\tfrac{1}{N}\Phi' W W' \Phi\right) \left(\tfrac{1}{N}\Phi'\Phi\right)^{-1} \\
&\to_{a.s.} \sigma^2_\varepsilon Q_\Phi(\nu)^{-1} + \sigma^2_\gamma Q_\Phi(\nu)^{-1} Q_{\Phi W}(\nu) Q_\Phi(\nu)^{-1},
\end{aligned}
$$
using Assumption~\ref{ass:qphiw} in the last step.
\end{proof}

\begin{proof}[Proof of Theorem~\ref{thm:parboot}]
From Theorem~\ref{thm:naivefail} it follows that:
$$
\sqrt{N}(\hat\beta^{b} - \hat\beta) \to_{d} \mathcal{N}\left(0,\ \hat\sigma^2_\varepsilon Q_\Phi(\nu)^{-1} + \hat\sigma^2_\gamma Q_\Phi(\nu)^{-1} Q_{\Phi W}(\nu) Q_\Phi(\nu)^{-1}\right),
$$
which is equal to $\sqrt{N}(\hat\beta - \beta)$ asymptotically as soon as $\hat\sigma_\gamma$ and $\hat\sigma_\varepsilon$ are consistent. We now show that they are. Since $\hat u$ are uniformly consistent estimates for $u$ as long as $\hat\beta$ is consistent, $\hat\varepsilon$ converge uniformly to $\tilde\varepsilon$, the residuals from the regression of the true $u$ on $W$. Denote $P_W$ and $M_W$ the projection on $\operatorname{span}(W)$ and the residual from that projection, respectively. Note that both these matrices are symmetric, meaning that $X^{T} = X$, and idempotent, meaning that $X^2 = X$. Finally, as $u = W\gamma + \varepsilon$, $M_W u = \underbrace{M_W W}_{=0}\gamma + M_W \varepsilon$. Hence:
$$
\frac{1}{N}\hat\varepsilon^{T}\hat\varepsilon \to_{a.s.} \frac{1}{N}\tilde\varepsilon^{T}\tilde\varepsilon = \frac{1}{N} u^{T} M_W M_W^{T} u = \frac{1}{N}\varepsilon^{T} M_W \varepsilon = \frac{1}{N}\varepsilon^{T}\varepsilon - \frac{1}{N}\varepsilon^{T} P_W \varepsilon.
$$
The first term converges almost surely to $\sigma^2_\varepsilon$, while the second term is a random variable with the variance equal to:
$$
2\Tr\left(P_W^{T} P_W\right) = 2\Tr(P_W) = 2\Tr\left(W(W^{T}W)^{-1}W^{T}\right) = 2\Tr\left(W^{T}W(W^{T}W)^{-1}\right) = 2K.
$$
As $K/N \to 0$, $\frac{1}{N}\hat\varepsilon^{T}\hat\varepsilon \to_{a.s.} \sigma^2_\varepsilon$. Finally, as $\frac{1}{N}\hat u^{T}\hat u \to_{a.s.} \sigma^2_\varepsilon + \sigma^2_\gamma N^{-1}\Tr(WW^{T})$, provided $N^{-1}\Tr(WW^{T})$ converges, $\hat\sigma^2_\gamma$ is also consistent.
\end{proof}

\end{document}